\documentclass[10pt]{article}

\usepackage{amssymb}
\usepackage{graphicx}
\usepackage[numbers,compress]{natbib}
\begin{document}

	\title{Joos-Weinberg covariant field with mass and spin $\frac{3}{2}$}
	
	\author{Ion I. Cot\u aescu\thanks{Corresponding author E-mail:~~i.cotaescu@e-uvt.ro}\\
		{\it West University of Timi\c soara,} \\{\it V. Parvan Ave. 4,
			RO-300223 Timi\c soara}}
	
	\maketitle

\begin{abstract}
We present the explicit theory  of  the Joos-Weinberg covariant field with spin $\frac{3}{2}$ which is a eight-dimensional massive covariant field   transforming according to the representation $(\frac{3}{2},0)\oplus(0, \frac{3}{2})$  of the group $SL(2,\mathbb{C})$. As the transformation matrices of this representation are still unknown, we apply a new method for deriving them using exclusively maximally reducible representations, e. g.    $(1,0)\otimes(\frac{1}{2},0)$ instead of the irreducible one $(1,\frac{1}{2})=(1,0)\otimes(0,\frac{1}{2})$ we meet in  usual frameworks. After applying this method,  we obtain a  $12$-component  covariant field transforming according to the representation $[(1,0)\otimes(\frac{1}{2},0)]\oplus [(0,1)\otimes(0, \frac{1}{2})]$ which is  maximally reducible,  up to subspaces of irreducible representations of  the $SU(2)$ group. Consequently, after developing the theory in  the direct product basis of the representation $(1,0)\otimes(\frac{1}{2},0)$, we can separate the sector of spin half  revealing thus the genuine   Joos-Weinberg covariant  massive field of spin $\frac{3}{2}$. The theory of this field  is build naturally deriving the  field equation and associated matrices, Lagrangian formalism,  inner product and the  closed  expressions of the orthonormal mode spinors.

Pacs: 03.65.Pm
\end{abstract}

\section{Introduction}

 The first covariant field with spin $\frac{3}{2}$  which was studied  is  the Rarita-Schwinger \cite{RS} vector-spinor field which satisfies a first order field equation analogous to Dirac's one. The  vector-spinor representation  mixes components of spin $\frac{3}{2}$ and $\frac{1}{2}$ that must be separated imposing subsidiary constraints. In an external electromagnetic field these conditions can be violated for avoiding the risk of unwanted superluminal propagation modes \cite{VZ}. Fortunately, these problems have a natural solution in supergravity where the Rarita-Schwinger equation is considered now as the standard field equation of the gravitino, the supersymmetric partner of the graviton.
 
 However, we remain with the barionic resonances with spin $s>\frac{1}{2}$ for which we do not have yet explicit equations. 	
In Minkowski space-time the particles with spin half or higher are studied from long time applying the theory of  finite-dimensional representations $(j_1,j_2)=(j_1,0)\otimes(0,j_2)$ of  the $SL(2,\mathbb{C})$ group  that  can be  manipulated formally as, for example, in Ref. \cite{Barut} for extracting invariant amplitudes.  A more specific framework was proposed by  Joos \cite{JW1} and Weinberg \cite{JW2} which assumed that the massive covariant fields with single spin $s$ must transform according to the reducible  representations $(s,0)\oplus (0,s)$. In what follows we say that these are  Joos-Weinberg covariant fields with spin $s$, denoted from now as $JW(s)$, understanding that here we consider massive fields only. Note that the  $JW(\frac{1}{2})$ field is just the Dirac one. 

The principal pieces of the Joos-Weinberg approach are the Wigner boosts  that define the form of covariant fields and, implicitly, the structure of field equations and propagators one needs to formulate  the Feynman rules for any spin \cite{JW2}. Unfortunately,  the transformation matrices of the representations $(s,0)\oplus (0,s)$,  in concrete bases of their carrier spaces,  were never written explicitly for $s>1$ because of some technical difficulties  in computing the series of  the exponential mappings, $\exp : sl(2,\mathbb{C}) \to SL(2,\mathbb{C})$, involving $sl(2,\mathbb{C})$ generators which do not have diagonal form in the basis  under consideration in the carrier space.   

 For this reason,  one resorts to alternative methods of  constructing larger representations, considering  either direct products  of  Dirac spinors   \cite{C1,C2,C3} or tensor-spinor  products  from which single spin representations have to be extracted using  systems of projection operators \cite{A,A1,A2,A3,A4}.  This method is laborious because of the irreducible representations $(j_1,j_2)$ with $j_1\not=0$ and $j_2\not=0$ whose subspeces of spin $s=j_1+j_2,j_1+j_2-1,\cdots |j_1-j_2|$ cannot be separated as long as the boost transformations are not reducible, mixing different spin components. The typical  example is  the Rarita-Schwinger field with spin $\frac{3}{2}$ build in a  frameworks including the reducible representation $(1,\frac{1}{2})\oplus(1,\frac{1}{2})$ whose subspaces of spin $\frac{3}{2}$ and $\frac{1}{2}$ cannot be separated without imposing the aforementioned constraints.  Other examples are discussed in Ref. \cite{B}.
 
 For overcoming the technical difficulties in evaluating exponential series we would like to propose here a new framework involving exclusively  maximally reducible representations. The idea is to consider only direct products of single  spin representations of  same type, either $(j,0)$ or $(0,j)$, without mixing  them in the same direct product. This can be done at any time considering the reducible representations $(j_1,0)\otimes(j_2,0)$ instead of  the irreducible ones,  $(j_1,j_2)=(j_1,0)\otimes(0, j_2)$.  For example, the irreducible representation $(1,\frac{1}{2})$ may be replaced by the direct product $(1,0)\otimes(\frac{1}{2},0)$ for obtaining a maximally reducible representation $(\frac{3}{2},0)\oplus (\frac{1}{2},0)$. 

Our principal goal in this paper  is to apply the method of maximally reducible representations for deriving explicitly the  transformation matrices of Wigner boosts in canonical rotation bases of the representation $(\frac{3}{2},0)\oplus (0,\frac{3}{2})$.  These matrices have to be used  for
constructing the field equation and the orthonormal mode spinors of the $8$-dimensional covariant field of spin $\frac{3}{2}$ in chiral formalism   \cite{A}.  In other words, we would like to obtain the explicit forms of  field equation and mode spinors of the $JW(\frac{3}{2})$ field.

For this purpose we start with a $12$- components covariant field whose  left-handed components transform according to the maximally reducible representation $(1,0)\otimes(\frac{1}{2},0)$ while the right-handed ones according to the adjoint representation $(0,1)\otimes(0,\frac{1}{2})$. 
Applying then the Joos-Weinberg method, we construct the field equations in momentum representation  assuming that in rest frames these are proportional to the metric matrix which is in the same time parity operator \cite{A}. Furthermore,  we develop the theory of  resulted 12-dimensional spinor field first in the direct product basis of mentioned representations deriving the field equations in momentum representation after outlining the Lagrangian formalism. We find thus that this field has two components, one of spin $\frac{1}{2}$ having a first order equation and another one of spin $\frac{3}{2}$  satisfying a third order equation \cite{Ham1,Ham2,A}. In the last case the Lagrangian density with second order derivatives \cite{N} leads to a new type of inner product involving first and second order derivatives. After obtaining the  field equations, we take advantage of our  maximally reducible approach allowing us to separate the spin sectors in canonical rotation basis. In this manner we find a field of spin $\frac{1}{2}$ which satisfies the Dirac equation and the desired $JW(\frac{3}{2}$  transforming according to the representation $(\frac{3}{2},0)\oplus(0, \frac{3}{2})$. Focusing on this last one, we study the matrices associated to the field equation, the mass condition and the closed expressions of  normalized mode spinors with physical meaning. 

This paper is organized as follows. In the next section we present briefly the finite-dimensional representations of the $SL(2,\mathbb{C})$ group and the general structure of covariant spinor fields in chiral formalism \cite{A}. The next section is devoted to the maximally reducible representations in direct product basis.  In the first subsection we define the direct product basis showing how this is related to the canonical rotation basis and giving the projection operators. In the next subsection we present the Lagrangian formalism with first and second order derivatives, leading to a first order equation for the components of spin half and a third order one for the components of spin $\frac{3}{2}$. Moreover, we show how the conserved current density suggests the form of specific inner product of fields with spin $\frac{3}{2}$. In the third subsection we derive the form of field equations in  momentum representation  using the direct product basis.  We show that the field equation depend on $\gamma$-matrices formed by blocks of $\sigma$-matrices which play the same role as Pauli's matrices in Dirac's theory. The section 4 is devoted to the $JW(\frac{3}{2})$ 8-dimensional field.  In the first subsection we show how the spin sectors can be separated in canonical rotation basis where the field of spin half is precisely the Dirac spinor. In the next subsection we focus on the field  $JW(\frac{3}{2})$ studying its equation in momentum representation for deriving the principal associated matrices and outlining some specific properties. In the last subsection we derive the normalized mode spinors of this field using  the inner product defined previously for assuring the orthonormalization of the basis of mode spinors of spin $\frac{3}{2}$ with physical meaning. Moreover, we specify that there are infinite norm solutions which are nonphysical. In the last section we present some concluding remarks and summarize our method of minimally reducible representations.

In  Appendix A we show how the boosts are build using Kronecker's product while in B we give the new $\sigma$-matrices of the spin $\frac{3}{2}$ sector.

\section{Covariant fields}

The covariant fields of special relativity \cite{BDR} are defined on  Minkowski's space-time ${M}$ with  the metric tensor $\eta={\rm diag}(1,-1,-1,-1)$ and  the chart with Cartesian coordinates $x^{\mu}$ labeled by Greek indices  ($\alpha,\,\beta,...\mu,\,\nu...=0,1,2,3$). The fields $\psi$ transform covariantly under  Poincar\' e  isometries, $(\Lambda,a): x\to x'=\Lambda x+a$, that form the  group  $P_{+}^{\uparrow} ={T}(4)\,\circledS\, {L}_{+}^{\uparrow}$ \cite{WKT} constituted by the transformations $\Lambda \in {L}_{+}^{\uparrow}$ of the  orthochronous proper Lorentz group,  preserving the metric $\eta$, and the four dimensional translations, $a\in  \mathbb{R}^4$ of the invariant  subgroup $T(4)$. 

\subsection{Finite-dimensional representations of the $SL(2,\mathbb{C})$ group}

For  fields with half integer spins one considers, in addition,  the universal covering group of the Poincar\' e one,  $ \bar{ P}^{\uparrow}_{+}={T}(4)\,\circledS\, SL(2,\mathbb{C})$, formed by the mentioned translations and transformations  
	\begin{equation}\label{tr}
	\lambda(\omega)=\exp\left(-\frac{i}{2}\omega^{\alpha\beta}s_{\alpha\beta}\right)\in SL(2,\mathbb{C})\,, 
\end{equation}
depending on  real-valued parameters, $\omega^{\alpha \beta}=-\omega^{\beta\alpha}$
and the generators $s_{\alpha\beta}$ of the $sl(2, \mathbb{C})$ Lie algebra, re-denoted often as
\begin{equation}
		s_i= \frac{1}{2}\varepsilon_{ijk}s^{jk} \,, \quad k_i=s_{i0}\,, \quad (i,j,k...=1,2,3)\,,
\end{equation}
 when one needs to study separately the rotations $r\in SU(2)\subset SL(2,\mathbb{C}$ or the Lorentz boosts $l\in SL(2,\mathbb{C}/SU(2)$.
 
 We remind the reader that the operators $a_i=\frac{1}{2}(s_i+ik_i)$ and   $b_i=\frac{1}{2}(s_i-ik_i)$ form two independent $su(2)$ algebras such that any irreducible representation  $(j_a,j_b)$ of the $SL(2,\mathbb{C})$ group may be realized in  spaces ${\cal V}_{j_a}\otimes {\cal V}_{j_b}$ formed as direct products between spaces of unitary irreducible representations, $(j_a)$ and $(j_b)$, of the $su(2)$ algebras $\{a_i\}$ and respectively $\{b_i\}$ \cite{WKT,W}.  A special role play the single spin irreducible representations, $\tau_s=(s,0)$ and their conjugated, $\bar\tau _s=(0,s)$, whose generators 
 \begin{eqnarray}
 \tau_s(b_i)=0~~&\Rightarrow&~~\tau_s(s_i)=I_i^s\,, ~~~\tau_s(k_i)=-i I_i^s\,,\\
 \tau_s(a_i)=0~~&\Rightarrow&~~	\bar\tau_s(s_i)=I_i^s\,, ~~~\bar\tau_s(k_i)=i I_i^s\,,
 \end{eqnarray}
 can be expressed in terms of generators $I^s_i$ of  unitary irreducible representations $(s)$ of the $SU(2)$ group.  Once we have such irreducible representations  we can construct any  representation $\tau=(j_1,j_2)$ as the direct product $\tau=\tau_{j_1}\otimes\tau_{j_2}$.  
 Given an irreducible representation  $\tau=(j_1,j_2)$ we say that   $\bar \tau=(j_2,j_1)$ is its conjugated or simply adjoint representation.   When $j_1=j_2$ the irreducible representation $\tau=\bar\tau$ is selfadjoint. For $j_1\not=j_2$ the  irreducible representations $\tau$ and $\bar\tau$ are not equivalent being related as 
 \begin{eqnarray}
 	\bar\tau(s_i)=\tau(s_i) &\Rightarrow&\bar\tau(r)=\tau(r)\,, ~~~~\forall r\in SU(2)\\ 	
 \bar\tau(k_i)=-\tau(k_i) &\Rightarrow&\bar\tau(l)=\tau(l^{-1})\,,~\forall l\in SL(2,\mathbb{C}) /SU(2).\label{t6}
 \end{eqnarray}
 Moreover, these  representations have the Hermitian properties
 \begin{eqnarray}
 	\tau(s_i)^+=\tau(s_i) &\Rightarrow&\tau(r)^+=\tau(r^{-1})\,, ~~~~\forall r\in SU(2)\\ 	
 	\tau(k_i)^+=-\tau(k_i) &\Rightarrow&\tau(l)^+=\tau(l)\,,~\forall l\in SL(2,\mathbb{C}) /SU(2).\label{t8}
 \end{eqnarray}
 that can be gathered in compact form,
 \begin{equation}\label{tt}
 	\tau(\lambda)^+=\bar\tau(\lambda^{-1})\,,\quad \forall \lambda\in SL(2,\mathbb{C})\,.
 \end{equation}
As the irreducible representations $\tau$ and $\bar\tau$  have the same spin content, $s=j_1+j_2, j_1+j_2-1...|j_1-j_2|$ we may use similar bases, either 
the {\em direct product basis}, 
\begin{equation}
	|j_1,\mu_1\rangle\otimes|j_2,\mu_2\rangle \,, \quad \begin{array}{l}
		\mu_1=j_1,j_1-1,...-j_1\\
		 \mu_2=j_2,j_2-1,...-j_2
		 \end{array} \,,
\end{equation}
or the canonical rotation basis whose vectors 
  \begin{equation}\label{CG}
 	|\tau,s,\sigma\rangle=\sum_{\mu_1+\mu_2=\sigma}|j_1,\mu_1\rangle\otimes|j_2,\mu_2\rangle \langle j_1,\mu_1;j_2,\mu_2|s,\sigma\rangle \,,
 \end{equation}
are derived with the help of the Clebsh-Gordan coefficients \cite{WKT}. In what follows we say simply that this is the {\em canonical basis}.

The  vector irreducible representation $\tau_V=(\frac{1}{2},\frac{1}{2})$ defines the Lorentz group  ${L}_{+}^{\uparrow}$ for which we use the traditional notations  with capital symbols,  $\tau_V(\lambda) = \Lambda(\lambda)\in {L}_{+}^{\uparrow}$.   The representation space ${\cal V}_V$ is formed by the vectors with contravariant indices transforming as $V^{\mu}\to \Lambda ^{\mu\,\cdot}_{\cdot\,\nu}V^{\nu}$. As these vectors may take complex values,  it is convenient to use the matrix  formalism considering  $V$ as a column  of components $V^{\mu}$	whose adjoint $\bar{V}=V^+\eta$ is the line-matrix of components $V_{\mu}^*$. With these notations the invariant Hermitian forms may be written as $\eta_{\mu\nu}V^{\mu\,*}V^{\prime\,\nu}=V^+\eta V'=\bar{V} V'$. Particularly, the mass-shell condition reds now $\bar p p-m^2=0$ instead of $p^2-m^2=0$.

\subsection{Covariant fields with half integer spin}

In this framework the covariant fields with spin, $\psi:M\to {\cal V}_{\rho}$, may be defined on $M$ with values in vector spaces ${\cal V}_{\rho}$ carrying  reducible finite-dimensional representations  $\rho$  of the   $SL(2,\mathbb{C}) $ group where invariant Hermitian  forms can be defined. This exigence can be accomplished by choosing reducible representations  including pairs of adjoint irreducible representations $\tau\oplus \bar\tau =(J_1,j_2)\oplus (j_2,j_1)$, as well as selfadjoint irreducible representations, $\tau=\bar\tau=(j,j)$ but which appear only in the case of integer spins. As here we study a field with half integer spin we may  resort to the chiral formalism considering the reducible representation $\rho=\rho_L\oplus\rho_R$ formed by the left (L) and right (R) handed  chiral components assumed to be adjoint with each other,
\begin{equation}
	\rho_L=\tau\oplus\tau'\oplus...  ~~\Rightarrow~~ \rho_R=\bar\rho_L=\bar\tau\oplus\bar\tau'\oplus...
	\end{equation} 
 The space ${\cal V}_{\rho}={\cal V}_L\oplus {\cal V}_R$ is the orthogonal sum of the spaces carrying the representations $\rho_L$ and $\rho_R$ such that the covariant fields  can be represented as
\begin{eqnarray}
	\psi(x)=\psi_L(x) + \psi_R(x)= 	\left( \begin{array}{c}
		\phi_L(x)\\
		\phi_R(x)
\end{array}\right) \,,~~~\begin{array}{c} \phi_L\in{\cal V}_L\,,\\\phi_R\in{\cal V}_R\,.\end{array}
\end{eqnarray}
The fields $\psi$ transform   under isometries  according to the {covariant} representation  ${T}   \,:\,(\lambda, a)\to { T}_{\lambda,a}\in {\rm Aut}({\cal V}_{\rho})$ of the group $ \tilde{P}^{\uparrow}_{+}$ which has the action \cite{WKT}
\begin{eqnarray}
		({T}_{\lambda,a}\psi  )(x)
		=\rho(\lambda) \psi  \left(\Lambda(\lambda)^{-1}(x-a)\right)
		=\left(\begin{array}{cc}
			\rho_L(\lambda)&0\\
			0&\rho_R(\lambda)
		\end{array}\right) 	\left( \begin{array}{c}
		\phi_L(\Lambda(\lambda)^{-1}(x-a))\\
		\phi_R(\Lambda(\lambda)^{-1}(x-a))
		\end{array}\right)\,.\label{TAa}
	\end{eqnarray}
Working with many kind of matrices it is useful to denote by ${\frak M}(n)$ the algebra of $n\times n$-matrices with complex-valued matrix elements.  Taking   $n={\rm dim}(\rho_L)={\rm dim}(\rho_R)$ we have
\begin{equation}
	\rho_L,\,\rho_R\in{\frak M}(n)~~\Rightarrow~~\rho=\rho_L\oplus\rho_R \,\in\,{\frak M}(2n)\,,
\end{equation}  
avoiding redundant explanations. In this lay out any matrix is formed by chiral blocks
\begin{equation}
	M=\left( \begin{array}{cc}
		M_L&M_{LR}\\
	M_{RL}&M_R
	\end{array}\right)\, \in {\frak M}(2n)\,,\quad M_L,M_{LR},...\in {\frak M}(n)\,,
\end{equation}
which may be considered separately.

The diagonal blocks of $\rho(\lambda)$ have the remarkable property
\begin{equation}
	\rho_{L/R}^+(\lambda)=\rho_{R/L}(\lambda^{-1})\,,
\end{equation} 	
resulted from the rule (\ref{tt}) of  adjoint irreducible representations.
Therefore,  the scalar  $\phi_L^+\phi_R+\phi_R^+\phi_L$ is invariant under transformations (\ref{TAa}). This expression can be put in compact form as $\overline{\psi}\psi$  defining the adjoint $\overline{\psi}=\psi^+\gamma$ of $\psi$ with the help of the new matrix 
\begin{equation}\label{gam}
	\gamma=\left( \begin{array}{cc}
		0&{\bf 1}\\
		{\bf 1}&0
	\end{array}\right)\in {\frak M}(2n)\,, \quad {\bf 1}=1_{n\times n}\in {\frak M}(n)  \,,
\end{equation}
which play the role of metric operator.  We say that the matrix $\overline\mu=\gamma\mu^+\gamma$ is the adjoint of the matrix $\mu$ which means that a selfadjoint matrix satisfies $\overline\mu=\mu$. As we may write
\begin{equation}
	\overline\rho(\lambda)=\gamma\rho^+(\lambda)\gamma=\rho(\lambda^{-1})\,,\quad
	\overline\rho(s_{\mu\nu})=\rho(s_{\mu\nu})\,,
\end{equation}
we say that the representations $\rho$ are pseudo-unitary having selfadjoint generators.
In addition, we consider the identity matrix $1_{\rho}$ of the representation $\rho$, the chiral one $\gamma^5$ and the chiral projection operators defined as
\begin{equation}\label{gam5}
	1_{\rho}=\left( \begin{array}{cc}
		{\bf 1}&0\\
		0&{\bf 1}
	\end{array}\right)\,, ~~
	\gamma^5=\left( \begin{array}{cc}
		-{\bf 1}&0\\
		0&{\bf 1}
	\end{array}\right)\,, ~~ P_{L/R}=\frac{1}{2}\left(1_{\rho}\mp\gamma^5 \right) \,.
\end{equation}
Obviously, we have $\gamma^2=(\gamma^5)^2=1_{\rho}$,  $\{\gamma,\gamma^5\}=0$, as well as
$\overline\gamma=\gamma$ and $\overline{\gamma^5}=-\gamma^5$.

\section{Twelve-component field with spins $\frac{3}{2}$ and $\frac{1}{2}$}

We now  tray to construct  the theory of massive particles of spin $\frac{3}{2}$ choosing simplest maximally reducible representations, 
\begin{eqnarray}
	\hat\rho_L&=&\textstyle{\tau_{1}\otimes \tau_{\frac{1}{2}}}
	\equiv\textstyle{(1,0)\otimes(\frac{1}{2},0)}\,,\label{RL}\\
	\hat\rho_R&=&\textstyle{\bar\tau_{1}\otimes \bar\tau_{\frac{1}{2}}}
	  \equiv\textstyle{(0,1)\otimes(0,\frac{1}{2})}\,,\label{RR}
\end{eqnarray} 
where $\hat\rho_L$,\, $\hat\rho_R\in{\frak M}(6)$  while $\hat\rho=\hat\rho_L\oplus\hat\rho_R \in {\frak M}(12)$. The representations $\hat\rho_L$ and $\hat\rho_R$ have the same spin content such that we may consider the same basis in the representation spaces ${\cal V}_L$ and ${\cal V}_R$.

\subsection{Direct product and canonical  bases}

  The first choice is the basis formed as the direct product between the canonical  bases of the irreducible representations $\tau_1$ and $\tau_{\frac{1}{2}}$,  $\{|1,\mu\rangle; \mu=1,0,-1\}$ and respectively $\{|\textstyle{\frac{1}{2}},\nu\rangle; \nu=\pm\frac{1}{2}\}$. Then the direct product basis can be realized as
\begin{eqnarray}
	|1,1\rangle\otimes|\textstyle{\frac{1}{2},\frac{1}{2}}\rangle&=&(1,0,0,0,0,0)^T=|\epsilon_1\rangle\,, \nonumber\\
	|1,1\rangle\otimes|\textstyle{\frac{1}{2},-\frac{1}{2}}\rangle&=&(0,1,0,0,0,0)^T=|\epsilon_2\rangle\,, \nonumber\\
	|1,0\rangle\otimes|\textstyle{\frac{1}{2},\frac{1}{2}}\rangle&=&(0,0,1,0,0,0)^T=|\epsilon_3\rangle\,, \nonumber\\
	|1,0\rangle\otimes|\textstyle{\frac{1}{2},-\frac{1}{2}}\rangle&=&(0,0,0,1,0,0)^T=|\epsilon_4\rangle\,, \nonumber\\
		|1,-1\rangle\otimes|\textstyle{\frac{1}{2},\frac{1}{2}}\rangle&=&(0,0,0,0,1,0)^T=|\epsilon_5\rangle\,, \nonumber\\
	|1,-1\rangle\otimes|\textstyle{\frac{1}{2},-\frac{1}{2}}\rangle&=&(0,0,0,0,0,1)^T=|\epsilon_6\rangle\,.\label{basis1}
\end{eqnarray}
In this basis, denoted by $\{\epsilon\}$, the chiral spinors  $\phi_L$ and $\phi_R$ are represented by their components that form column matrices,  as for example
\begin{equation}
\phi_L=\sum_{i=1}^6 |\epsilon_i\rangle\langle\epsilon_i|\phi_L\rangle=(\phi_L^1,\phi_L^2,...)^T	\,,\quad
\phi_L^i=\langle\epsilon_i|\phi_L\rangle\,,
\end{equation}
  and similarly for $\phi_R$. The matrices $M\in{\frak M}(6)$ have matrix elements 
  $M_{ij}=\langle\epsilon_i|M|\epsilon_j\rangle$. In what follows  we prefer the simple matrix notation instead of the bra-ket formalism but which is indispensable for understanding the basic definitions. We must specify that  in the 12-dimensional space ${\cal V}={\cal V}_L\oplus {\cal V}_R$ we may consider the basis $\{\epsilon\}_L\oplus \{\epsilon\}_R$ introducing a supplemental index indicating the chirality. For avoiding excessive  notations we tray to study the algebraic properties of chiral blocks considered separately.
  
  Starting  with the basis (\ref{basis1})  we have  the opportunity of  constructing the transformation matrices using the direct (or Kronecker) products,
\begin{equation}
	\hat\rho_L(\lambda)=\tau_1(\lambda)\otimes\tau_{\frac{1}{2}}(\lambda)\,,\quad
	\hat\rho_R(\lambda)=\bar\tau_1(\lambda)\otimes\bar\tau_{\frac{1}{2}}(\lambda)\,,
\end{equation} 
involving simpler matrices we know. In Appendix A we show how the boosts $\hat\rho_L(l_{\bf p})$ and $\hat\rho_R(l_{\bf p})$ may be derived in this manner for obtaining the final result (\ref{T1}).

However, in the basis   $\{\epsilon\}$ we cannot separate the components of spin $\frac{3}{2}$ from those of spin  $\frac{1}{2}$ such that  we need to resort to the canonical  basis (\ref{CG}) that can be obtained using the Clebsh-Gordan coefficients listed in Ref. \cite{WKT}. We obtain thus the canonical basis of spin $\frac{3}{2}$, 
\begin{eqnarray}
	|\textstyle{\frac{3}{2},\frac{3}{2}}\rangle&=&\sum_i|\epsilon_i\rangle\langle\epsilon_i	|\textstyle{\frac{3}{2},\frac{3}{2}}\rangle=(1,0,0,0,0,0)^T=\xi_{\textstyle{\frac{3}{2},\frac{3}{2}}}\,, \nonumber\\
		|\textstyle{\frac{3}{2},\frac{1}{2}}\rangle&=&\sum_i|\epsilon_i\rangle\langle\epsilon_i	|\textstyle{\frac{3}{2},\frac{1}{2}}\rangle=(0,a,b,0,0,0)^T=\xi_{\textstyle{\frac{3}{2},\frac{1}{2}}}\,, \nonumber\\
		|\textstyle{\frac{3}{2},-\frac{1}{2}}\rangle&=&\sum_i|\epsilon_i\rangle\langle\epsilon_i		|\textstyle{\frac{3}{2},-\frac{1}{2}}\rangle=(0,0,0,b,a,0)^T=\xi_{\textstyle{\frac{3}{2},-\frac{1}{2}}}\,, \nonumber\\
		|\textstyle{\frac{3}{2},-\frac{3}{2}}\rangle&=&\sum_i|\epsilon_i\rangle\langle\epsilon_i		|\textstyle{\frac{3}{2},-\frac{3}{2}}\rangle=(0,0,0,0,0,1)^T=\xi_{\textstyle{\frac{3}{2},-\frac{3}{2}}}\,, \label{basis2a}~~~~~
\end{eqnarray}
where $a= \frac{1}{\sqrt{3}}$ and   $b=\sqrt{\frac{2}{3}}$. Similarly, for $s=\frac{1}{2}$ we have
\begin{eqnarray}
	|\textstyle{\frac{1}{2},\frac{1}{2}}\rangle&=&\sum_i|\epsilon_i\rangle\langle\epsilon_i		|\textstyle{\frac{1}{2},\frac{1}{2}}\rangle=(0,b,-a,0,0,0)^T=\xi_{\textstyle{\frac{1}{2},\frac{1}{2}}}\,, \nonumber\\
|\textstyle{\frac{1}{2},-\frac{1}{2}}\rangle&=&\sum_i|\epsilon_i\rangle\langle\epsilon_i		|\textstyle{\frac{1}{2},-\frac{1}{2}}\rangle=(0,0,0,a,-b,0)^T=\xi_{\textstyle{\frac{1}{2},-\frac{1}{2}}}\,, \label{xi1}\label{basis2b}~~~~~~
\end{eqnarray}
The spinors $\xi_{s,\sigma}$ form orthonormal bases, $\xi^+_{s,\sigma}	\xi_{s',\sigma'}=\delta_{ss'}\delta_{\sigma\sigma'}$,  giving rise to  projection operators, 
\begin{equation}
	\sum_{\sigma=-s}^{s}\xi_{s,\sigma}\xi^+_{s,\sigma}=\pi_{s}\in{\frak M}(6)\,,\quad s=\textstyle{\frac{3}{2}, \,\frac{1}{2}}\,,
\end{equation}
on the subspaces of spin $s$ of ${\cal V}_L$ or ${\cal V}_R$. In direct product basis $\{\epsilon\}$ these projection operators read   
\begin{equation}
\pi_{\frac{3}{2}}=\left(\begin{array}{cccccc}
	1&0&0&0&0&0\\
	0&\frac{1}{3}&\frac{\sqrt{2}}{3}&0&0&0\\	 
	0&\frac{\sqrt{2}}{3}&\frac{2}{3}&0&0&0\\
	0&0&0&\frac{2}{3}&\frac{\sqrt{2}}{3}&0\\
	0&0&0&\frac{\sqrt{2}}{3}&\frac{1}{3}&0\\
	0&0&0&0&0&1
		\end{array} \right)\,, \qquad \pi_{\frac{1}{2}}={1}_{6\times6}-\pi_{\frac{3}{2}}\,.
	\end{equation}
Using these matrices we construct the projection operators 
\begin{equation}
	P_s=\left(\begin{array}{cc}
		\pi_s&0\\
		0&\pi_s
		\end{array}\right)\in {\frak M}(12)\,, \quad s=\textstyle{\frac{3}{2},\,\frac{1}{2}}\,.
\end{equation}
which form a complete system of orthogonal projection operators splitting the space ${\cal V}_{\hat\rho}$ in subspaces of given spin. Any field $\psi\in{\cal V}_{\hat\rho}$ can be written as 
\begin{equation}
	\psi=P_{\frac{3}{2}}\psi+P_{\frac{1}{2}}\psi
	=\psi_{\frac{3}{2}} +\psi_{\frac{1}{2}}
\end{equation}
 while the matrices $X\in {\frak M}(12)$ may be decomposed in a similar manner,
 	\begin{equation}
 		X=P_{\frac{3}{2}}XP_{\frac{3}{2}}+P_{\frac{1}{2}}XP_{\frac{1}{2}}+P_{\frac{3}{2}}XP_{\frac{1}{2}}+P_{\frac{1}{2}}XP_{\frac{3}{2}}\,.
 	\end{equation}	
 When 	$P_{\frac{3}{2}}XP_{\frac{1}{2}}=P_{\frac{1}{2}}XP_{\frac{3}{2}}=0$  the matrix $X$ is maximally reducible  allowing the splitting 
 \begin{equation}
 	X=X_{\frac{3}{2}}+X_{\frac{1}{2}}\,,\qquad X_s=P_sXP_s=XP_s=P_sX\,.
 \end{equation} 
 Our approach based on the representation (\ref{RL}) and its adjoint (\ref{RR}) is maximally reducible, such that the matrices 
 \begin{equation}
 	\hat\rho(\lambda)=P_{\frac{3}{2}}\hat\rho(\lambda)+P_{\frac{1}{2}}\hat\rho(\lambda)=
 \hat\rho_{\frac{3}{2}}(\lambda)+\hat\rho_{\frac{1}{2}}(\lambda)	
 \end{equation}
 may be decomposed in terms of representations $\hat\rho_s$ which are equivalent to the representations $(s,0)\oplus(0,s)$ resulted after decomposing the initial direct products (\ref{RL}) and (\ref{RR}).

\subsection{Lagrangian formalism and inner products}

The field equations of our approach are expected to be of the first order for $s=\frac{1}{2}$ and third one for $s=\frac{3}{2}$, having the general form 
\begin{eqnarray}
\Gamma_{\frac{1}{2}}(i\partial_{\mu})\psi_{\frac{1}{2}}-m\,\psi_{\frac{1}{2}}=0\,,
~~&\quad& 	\Gamma_{\frac{1}{2}}(i\partial_{\mu})=i\hat\gamma^{\mu}\partial_{\mu}\label{eqD}\,,\\
\Gamma_{\frac{3}{2}}(i\partial_{\mu})\psi_{\frac{3}{2}}-m^3\psi_{\frac{3}{2}}=0\,, &\quad& 	\Gamma_{\frac{3}{2}}(i\partial_{\mu})=i^3\hat\gamma^{\mu\nu\sigma}\partial_{\mu}\partial_{\nu}\partial_{\sigma}\,,\label{eq}
\end{eqnarray}
where $\hat\gamma^{\mu}$ and $\hat\gamma^{\mu\nu\sigma}$ are matrices of ${\frak M}(12)$ supposed to be selfadjoint, $\overline{\hat\gamma^{\mu}}=\hat\gamma^{\mu}$ and $\overline{\hat\gamma^{\mu\nu\sigma}}=\hat\gamma^{\mu\nu\sigma}$,  while the last ones are symmetric in all their vector indices. The relativistic covariance of the entire theory requires these matrices to transform as
\begin{eqnarray}
	\hat\rho(\lambda^{-1})\hat\gamma^{\mu}\hat\rho(\lambda)&=&\Lambda^{\mu\,\cdot}_{\cdot\,\alpha}(\lambda)\hat\gamma^{\alpha}\label{rc1}\,,\\
	\hat\rho(\lambda^{-1})\hat\gamma^{\mu\nu\sigma}\hat\rho(\lambda)&=&\Lambda^{\mu\,\cdot}_{\cdot\,\alpha}(\lambda)\Lambda^{\nu\,\cdot}_{\cdot\,\beta}(\lambda)\Lambda^{\sigma\,\cdot}_{\cdot\,\gamma}(\lambda)\hat\gamma^{\alpha\beta\gamma}\,. \label{rc}
\end{eqnarray}
The spaces of solutions of the above equations have to be equipped with inner products that can be defined only in the context of  the Lagrangian formalism.

In what follows we discuss this formalism separately for the first and third order equations  omitting for simplicity the spin indices of the corresponding fields. We start with the field $\psi=\psi_{\frac{1}{2}}$ whose first order equation (\ref{eqD}) may be derived from the action 
\begin{equation}
		{\cal S}_{\frac{1}{2}}[\psi,\overline{\psi}]=\int d^4 x {\cal L}_{\frac{1}{2}}(\psi, \overline{\psi})
\end{equation} given by the well-known Lagrangian density 
\begin{equation}
{\cal L}_{\frac{1}{2}}(\psi, \overline{\psi})=\frac{i}{2}\left(\overline\psi\hat\gamma^{\mu}\psi_{,\mu}  -\overline\psi_{,\mu}\hat\gamma^{\mu}\psi  \right)	-m\overline\psi \psi\,.
\end{equation}
It is known that the conserved current density $j^{\mu}=\overline\psi\hat\gamma^{\mu}\psi$ induces the inner product 
\begin{equation}
	\langle \psi,\psi'\rangle_{\frac{1}{2}}=\int d^3x\, \overline\psi\hat\gamma^0\psi' \,,
\end{equation}
similar to that of Dirac's theory.

For $s=\frac{3}{2}$ we re-denote $\psi=\psi_{\frac{3}{2}}$ writing  Eq. (\ref{eq}) and its adjoint  in a more compact form, 
\begin{eqnarray}
	{E}[\psi]\equiv i\hat\gamma^{\mu\nu\sigma}\psi_{,\mu\nu\sigma}+m^3\psi=0\,, \label{eq1}\\ 
	\bar{ E}[\overline\psi]\equiv i\overline\psi_{,\mu\nu\sigma}\hat\gamma^{\mu\nu\sigma}-m^3\overline\psi=0\,.\label{eq2}
\end{eqnarray}
These equations can be derived  from the action 
\begin{equation}\label{actionD}
	{\cal S}_{\frac{3}{2}}[\psi,\overline{\psi}]=\int d^4 x {\cal L}_{\frac{3}{2}}(\psi, \overline{\psi})\,,
\end{equation}
whose Lagrangian density, 
\begin{eqnarray}
	{\cal L}_{\frac{3}{2}}(\psi, \overline{\psi})=N\left[\frac{i}{2}\left(  \overline\psi_{,\alpha\mu}\hat\gamma^{\alpha\mu\beta}\psi_{,\beta} -	\overline\psi_{,\alpha}\hat\gamma^{\alpha\mu\beta}\psi_{,\mu\beta} \right)+m^3\overline\psi \psi\right]\,,
\end{eqnarray}
with first and second order derivatives gives  the desired Euler-Lagrange equations, 
\begin{eqnarray}
	0=\frac{\partial{\cal L}}{\partial\overline\psi}-\partial_{\mu}\frac{\partial{\cal L}}{\partial\overline\psi_{,\mu}}+\partial_{\mu}\partial_{\nu}\frac{\partial{\cal L}}{\partial\overline\psi_{,\mu\nu}}={E}[\psi]\,,\\
	0=\frac{\partial{\cal L}}{\partial \psi}-\partial_{\mu}\frac{\partial{\cal L}}{\partial\psi_{,\mu}}+\partial_{\mu}\partial_{\nu}\frac{\partial{\cal L}}{\partial\psi_{,\mu\nu}}=-\bar{E}[\overline\psi]\,.	
\end{eqnarray}
The factor  $N\propto m^{-2}$, defined for the time being up to a numeric factor, is introduced for assuring  the action to remain dimensionless when the field has the usual dimension $[\psi]={\frac{3}{2}}$. The final value of $N$  will  be specified later.

Furthermore, applying the Noether theorem for the electromagnetic gauge, $\psi\to e^{-i\alpha}\psi\simeq\psi-i\alpha\psi+...$, we obtain the conserved current density \cite{N},
\begin{eqnarray}
	j^{\mu}(x)&=&i N\left[\left(\frac{\partial{\cal L}}{\partial\psi_{,\mu}}-\partial_{\nu}\frac{\partial{\cal L}}{\partial\psi_{,\mu\nu}} \right) \psi+ \frac{\partial{\cal L}}{\partial\psi_{,\mu\nu}}\psi_{,\nu}.-\overline\psi\left(\frac{\partial{\cal L}}{\partial\overline\psi_{,\mu}}-\partial_{\nu}\frac{\partial{\cal L}}{\partial\overline\psi_{,\mu\nu}} \right)-\overline\psi_{,\nu} \frac{\partial{\cal L}}{\partial\overline\psi_{,\mu\nu}}\right]\nonumber\\	
	&=&N\left[	\overline\psi_{,\alpha}(x)\hat\gamma^{\alpha\mu\beta}\psi_{,\beta}(x)-	\overline\psi(x)\hat\gamma^{\alpha\mu\beta}\psi_{,\alpha\beta}(x) -\overline\psi_{,\alpha\beta}(x)\hat\gamma^{\alpha\mu\beta}\psi(x)\right] \,,\label{cur}
\end{eqnarray}
which satisfies $\partial_{\mu}j^{\mu}(x)=0$ when $\psi$ and $\overline\psi$ are solutions of Eqs. (\ref{eq1}) and (\ref{eq2}). This inspire us to set the value of $N$ defining the relativistic inner product of this field as
\begin{eqnarray}
N=\frac{1}{3m^2}~~\Rightarrow~~\langle\psi,\psi'\rangle_{\frac{3}{2}}=\frac{1}{3 m^2}\int d^3 x\,\left(
	\overline\psi_{,\alpha}(x)\hat\gamma^{\alpha0\beta}\psi'_{,\beta}(x)-\overline\psi(x)\hat\gamma^{\alpha0\beta}\psi'_{,\alpha\beta}(x) -\overline\psi_{,\alpha\beta}(x)\hat\gamma^{\alpha0\beta}\psi'(x) \right)\,.\label{sp}
\end{eqnarray}
We choose  the factor $N=(3 m^2)^{-1}$ for compensating the existence of three terms giving similar contributions.

\subsection{Deriving field equations in momentum representation}

In momentum representation  all the quantities are defined on orbits in momentum space, $\Omega_{\mathring p}=\{ p|\, { p}=\Lambda \mathring{p}, \Lambda\in L_{+}^{\uparrow} \}$, that may be built by applying Lorentz transformations on a {representative} momentum $\mathring{p}$ \cite{Wig, Mc}.  In the case of massive particles, we discuss here,  the representative momentum is just the rest frame one,  $\mathring{p}=(m,0,0,0)$.  This can be transformed in an arbitrary  mass-shell momentum,  $p=(E_p,{\bf  p})\in \Omega_{\mathring{p}}$, of energy $E_p=\sqrt{{\bf p}^2+m^2}$, with the help of the  transformation  ${p}=\Lambda(l_{\bf  p})\,\mathring{p}$ where $l_{\bf p}\in SL(2,\mathbb{C})/SU(2)$ is a Lorentz boost parametrized as in Appendix A.

The general solutions $\psi$ of the field equation (\ref{eq}),  may be expanded in terms of  mode spinors,  $U_{{\bf  p},s\sigma}$ and  $V_{{\bf  p},s\sigma}$, of positive and respectively negative frequencies \cite{BDR},  
\begin{eqnarray}\label{Psi}
	\psi _s (x)=	\psi _s ^+(x)+	\psi_s  ^-(x)=\int d^3p \sum_{\sigma=-s}^{s}\left[U_{{\bf  p},{s}\,\sigma}(x) \alpha_{s\,\sigma}({\bf  p}) +V_{{\bf  p}, s\,\sigma}(x) \beta^ {*} _{s\, \sigma}({\bf  p})\right]\,,~~~~
\end{eqnarray}
where  $\alpha: \Omega_{\mathring p}\to {\cal V}$ and   $\beta: \Omega_{\mathring p}\to {\cal V}$ are  particle and respectively antiparticle { wave spinors} with values in the space ${\cal V}={\cal V}_{\frac{1}{2}}\oplus {\cal V}_{\frac{3}{2}}$ of  representation $(\frac{1}{2})\oplus(\frac{3}{2})$ of the $SU(2)$ group. The space of free fields ${\cal F}$  can be split  thus in two  subspaces of positive and respectively negative frequencies, ${\cal F}={\cal F}^+\oplus {\cal F}^-$, which must be  orthogonal with respect to the scalar product (\ref{sp}).

Looking for solutions of field equations, we observe that spinors of general form
\begin{eqnarray}
	U_{{\bf  p},s\sigma}(x)&=&u_{s\sigma}({\bf  p})\frac{1}{(2\pi)^{\frac{3}{2}}} \,e^{-iE_pt+i{\bf  p}\cdot{\bf  x}}\,,\label{U} \\
	V_{{\bf  p},s\sigma}(x)&=&v_{s\sigma}({\bf  p})\frac{1}{(2\pi)^{\frac{3}{2}}}\, e^{iE_pt-i{\bf  p}\cdot{\bf  x}}\,,\label{V}
\end{eqnarray} 
may be particular solutions of  Eqs. (\ref{eqD})  and  (\ref{eq}) only if  the spinors $u$ and $v$   satisfy  on mass-shell,
\begin{eqnarray}
	\Gamma_s(p) u_{s\sigma}({\bf p})-m^{2s}u_{s\sigma}({\bf p})&=&0\,,	~\\
	~~\Gamma_s(p) v_{s\sigma}({\bf p})+m^{2s}v_{s\sigma}({\bf p})&=&0\,,
\end{eqnarray}
where
\begin{equation}
\Gamma_{\frac{1}{2}}(p)=\hat\gamma^{\mu}p_{\mu}\,,\qquad\Gamma_{\frac{3}{2}}(p)=\hat\gamma^{\mu\nu\sigma}p_{\mu}p_{\nu}p_{\sigma}~~\in{\frak M}(12)\,.
\end{equation}
In the rest frame where $p=\mathring{p}$ we have
\begin{eqnarray}
	&&	\Gamma_{\frac{1}{2}}(\mathring{p})=m\hat\gamma^{0}\Rightarrow \left\{ \begin{array}{l}
		\hat\gamma^{0}u_{\frac{1}{2},\sigma}(0)=u_{{\frac{1}{2}}\sigma}(0)\,,\\
		\hat\gamma^{0}v_{\frac{1}{2},\sigma}(0)=-v_{{\frac{1}{2}},\sigma}(0)\,.
	\end{array}\right.\\
&&	\Gamma_{\frac{3}{2}}(\mathring{p})=m^3\hat\gamma^{000}\Rightarrow \left\{ \begin{array}{l}
			\hat\gamma^{000}u_{\frac{3}{2},\sigma}(0)=u_{\frac{3}{2},\sigma}(0)\,,\\
			\hat\gamma^{000}v_{\frac{3}{2},\sigma}(0)=-v_{\frac{3}{2},\sigma}(0)\,.
			\end{array}\right.
\end{eqnarray}
Hereby we understand that the field equations in momentum representation can be derived defining first the matrices $\hat\gamma^0$ and  $\hat\gamma^{000}$ as well as their eigenspinors, $u_{s\sigma}(0)$ and $v_{s\sigma}(0)$, applying then the rules  (\ref{rc1}) and  (\ref{rc}) which gives the final form of the matrices $\Gamma_s(p)$, 
\begin{eqnarray}
		m\hat\rho(l_{\bf p})\hat\gamma^{0}\hat\rho(l^{-1}_{\bf p})&=&m\Lambda^{0\,\cdot}_{\cdot\,\alpha}(l^{-1}_{\bf p})\hat\gamma^{\alpha}
	=\hat\gamma^{\alpha}p_{\alpha}=\Gamma_{\frac{1}{2}}(p)\\
	m^3\hat\rho(l_{\bf p})\hat\gamma^{000}\hat\rho(l^{-1}_{\bf p})&=&m^3\Lambda^{0\,\cdot}_{\cdot\,\alpha}(l^{-1}_{\bf p})\Lambda^{0\,\cdot}_{\cdot\,\beta}(l^{-1}_{\bf p})\Lambda^{0\,\cdot}_{\cdot\,\gamma}(l^{-1}_{\bf p})\hat\gamma^{\alpha\beta\gamma}\nonumber\\
	&=&\hat\gamma^{\alpha\beta\gamma}p_{\alpha}p_{\beta}p_{\gamma}=\Gamma_{\frac{3}{2}}(p)
\end{eqnarray}
after using the identity   $m\Lambda^{0\,\cdot}_{\cdot\,\mu}(l^{-1}_{\bf p})=p_{\mu}$ given by Eq, (\ref{Lal}).

For applying this method it is crucial to know the matrices $\hat\rho(l_{\bf p})\in {\frak M}(12)$. For this reason, following the steps indicated in Appendix A, we derive  the transformation matrices	in terms of null momentum components  (\ref{nul1}) as,  
		\begin{eqnarray}	
			\hat\rho_L(l_{\bf p})= \hat\rho_R(l^{-1}_{\bf p})=\tau_1(l_{\bf p})\otimes \tau_{\frac{1}{2}}(\l_{\bf p})
			=\frac{1}{(2m(E_p+m))^{\frac{3}{2}}}\times\hspace*{92mm}&&\nonumber\\
		{\footnotesize\left[ \begin {array}{cccccc}  \left( { E_-}+m \right) ^{3}&-
			\left( { E_-}+m \right) ^{2}{ p_-}&\sqrt {2} \left( { E_-}+m
			\right) ^{2}{ p_-}&\sqrt {2} \left( { E_-}+m \right) {{ p_-}}^{
				2}& \left( { E_-}+m \right) {{ p_-}}^{2}&-{{ p_-}}^{3}
			\\ \noalign{\medskip}- \left( { E_-}+m \right) ^{2}{ p_+}&D \left( 
			{ E_-}+m \right)  &\sqrt {2} \left( {
				E_-}+m \right) { p_-}{ p_+}&-\sqrt {2} Dp_- &-{{ p_-}}^{2}{ p_+}
			& \left( { E_+}+m \right) {{ p_-}}^{2}\\ \noalign{\medskip}-\sqrt 
			{2} \left( { E_-}+m \right) ^{2}{ p_+}&\sqrt {2} \left( { E_-}+m
			\right) { p_-}{ p_+}&C  \left( { E_-}+m \right) &- C { p_-}&-\sqrt {2} Dp_- &\sqrt {2} \left( {
				E_+}+m \right) {{ p_-}}^{2}\\ \noalign{\medskip}\sqrt {2} \left( 
			{ E_-}+m \right) {{ p_+}}^{2}&-\sqrt {2}Dp_+ &- C { p_+}& C  \left( { E_+}+m \right) &\sqrt 
			{2} \left( { E_+}+m \right) { p_-}{ p_+}&-\sqrt {2} \left( {
				E_+}+m \right) ^{2}{ p_-}\\ \noalign{\medskip} \left( { E_-}+m
			\right) {{ p_+}}^{2}&-{{ p_+}}^{2}{ p_-}&-\sqrt {2} Dp_+ &\sqrt {2}
			\left( { E_+}+m \right) { p_-}{ p_+}& D\left( { E_+}+m
			\right)  &- \left( { E_+}+m \right) ^
			{2}{ p_-}\\ \noalign{\medskip}-{{ p_+}}^{3}& \left( { E_+}+m
			\right) {{ p_+}}^{2}&\sqrt {2} \left( { E_+}+m \right) {{ p_+}}
			^{2}&-\sqrt {2} \left( { E_+}+m \right) ^{2}{ p_+}&- \left( { 
				E_+}+m \right) ^{2}{ p_+}& \left( { E_+}+m \right) ^{3}\end {array}
			\right] }\,, &&\nonumber\\
			\label{T1}
		\end{eqnarray}		
		using the notation
		\begin{eqnarray}
			C&=& 2E_+E_-+m(E_++E_-)\,, \nonumber   \\
			D&=&  (E_++m)(E_-+m)          \,.      \label{T1a}              
		\end{eqnarray}
In addition, we consider the matrices  \cite{A}   
\begin{equation}
\hat\gamma^0= P_{\frac{1}{2}}\hat\gamma=\left(\begin{array}{cc}
0&\pi_{\frac{1}{2}}\\
\pi_{\frac{1}{2}}&0
\end{array}\right)\,,\quad	\hat\gamma^{000}= P_{\frac{3}{2}}\hat\gamma=\left(\begin{array}{cc}
0&\pi_{\frac{3}{2}}\\
\pi_{\frac{3}{2}}&0
\end{array}\right)\,,
\end{equation}
where $\hat\gamma$  is  the matrix  ({\ref{gam}}) but now with ${\bf 1}=1_{6\times6}$. With this setting we may write  the following matrices on mass-shell,
\begin{eqnarray}\label{G1}
\left.	\Gamma_s(p)\right|_{\bar p p=m^2}=m^{2s}\hat\rho(l_{\bf p})P_s\hat\gamma\,\hat\rho(l_{\bf p}^{-1}) =\left(\begin{array}{cc}
	0&m^{2s} \hat\rho_L(l_{\bf p})\pi_s\hat\rho_L(l_{\bf p})\\ 
	m^{2s} \hat\rho_R(l_{\bf p})\pi_s\hat\rho_R(l_{\bf p})&0
	\end{array}\right)\,,
\end{eqnarray}
focusing then on their  blocks $LR$  that can be written off mass-shell substituting $E_p\to p^0$ and $m^2\to \bar pp$. We obtain  the matrices 
\begin{equation}
\Sigma_s(p)=\left.	m^{2s}\hat\rho_L(l_{\bf p})^2\pi_s\right|_{m^2\to \bar pp;\,E_p\to p^0}\,,
\end{equation}
 which represent the principal elements of our construction giving  the definitive expressions
\begin{equation}
	\Gamma_s(p)=\left(\begin{array}{cc}
		0&\Sigma_s(p)\\
		\bar\Sigma_s(p)&0\end{array}\right)\,,\quad \bar\Sigma_s(p^0,{\bf p})=\Sigma_s(p^0,-{\bf p})\,, \quad s=\textstyle{\frac{1}{2},\,\frac{3}{2}}\,,
\end{equation}
  that holds off mass-shell, for any $p=(p^0,\bf{p})\in \mathbb{R}^4_p$.

  These  matrices   give rise to any other algebraic component. With their  help   we may deduce  the form of $\hat\gamma$-matrices defining first the new matrices
  \begin{eqnarray}
  	\hat\sigma^{\mu}&=&\partial_{p_{\mu}}\Sigma_{\frac{1}{2}}(p) \,,\hspace*{22mm} 
  	\bar{\hat\sigma}^{\mu}=\partial_{p_{\mu}}\bar\Sigma_{\frac{1}{2}}(p)\,,\\
  	\hat\sigma^{\alpha\beta\gamma}&=&\frac{1}{3!}\partial_{p_{\alpha}}\partial_{p_{\beta}}\partial_{p_{\gamma}}\Sigma_{\frac{3}{2}}(p), ~~~~\,
  	\bar{\hat\sigma}^{\alpha\beta\gamma}=\frac{1}{3!}\partial_{p_{\alpha}}\partial_{p_{\beta}}\partial_{p_{\gamma}}\bar\Sigma_{\frac{3}{2}}(p)\,,\label{sig1}
  \end{eqnarray}
  which play the same role as the Pauli matrices in Dirac's theory. These allow us to construct the matrices
  \begin{eqnarray}
  	\hat\gamma^{\mu}=\left(\begin{array}{cc} 
  		0&\	\hat\sigma^{\mu}\\
  		\bar{\hat\sigma}^{\mu}&0
  	\end{array}\right) \,,\qquad
  	\hat\gamma^{\alpha\beta\gamma}=\left(\begin{array}{cc} 
  		0&\	\hat\sigma^{\alpha\beta\gamma}\\
  		\bar{\hat\sigma}^{\alpha\beta\gamma}&0
  	\end{array}\right)\,.\label{maga}
  \end{eqnarray}
  determining the form of  field equations  (\ref{eqD}) and (\ref{eq}).
  These matrices have the  following obvious properties,
  \begin{eqnarray}
  	(\hat\sigma^{\mu})^+&=&\hat\sigma^{\mu}\,,~~\qquad 
  	\bar{\hat\sigma}^{\mu}=\hat\sigma_{\mu}\,, ~~~~\qquad 
  	\overline{\hat\gamma^{\mu}}=\hat\gamma^{\mu}\,,\\	
  	(\hat\sigma^{\alpha\beta\gamma})^+&=&\hat\sigma^{\alpha\beta\gamma}\,,\quad 
  	\bar{\hat\sigma}^{\alpha\beta\gamma}=\hat\sigma_{\alpha\beta\gamma}\,, \quad 
  	\overline{\hat\gamma^{\alpha\beta\gamma}}=\hat\gamma^{\alpha\beta\gamma}\,,\label{sigma}
  \end{eqnarray}
  which assure the coherence of  entire theory of covariant fields.

  Following  the above procedure for $s=\frac{1}{2}$ we obtain  the matrix  $	\Sigma_{\frac{1}{2}}: \mathbb{R}^4_p\to {\frak M}(6)$ that reads
  \begin{equation}\label{Sighalf}
 	\Sigma_{\frac{1}{2}}(p)=\left(\begin{array}{cccccc}
 		0&0&0&0&0&0\\	
 		0&\frac{\sqrt{2}}{3}p^0_-&-\frac{\sqrt{2}}{3}p^0_-&-\frac{\sqrt{2}}{3}p_-&\frac{2}{3}p_-&0\\
 		0&-\frac{\sqrt{2}}{3}p^0_-&\frac{1}{3}p^0_-&\frac{1}{3}p_-&-\frac{\sqrt{2}}{3}p_-&0\\
 		0&-\frac{\sqrt{2}}{3}p_+&\frac{1}{3}p_+&\frac{1}{3}p^0_+&-\frac{\sqrt{2}}{3}p^0_+&0\\
 		0&\frac{2}{3}p_+&-\frac{\sqrt{2}}{3}p_+&-\frac{\sqrt{2}}{3}p^0_+&\frac{2}{3}p^0_+&0\\
 		0&0&0&0&0&0\\
 	\end{array} \right)\,, 
 \end{equation}
where now $p^0_{\pm}=p^0\pm p^3$  are  null momentum components  off mass-shell. The matrix $\Sigma_{\frac{3}{2}}(p)$ has a more complicated form that cannot be written here. We give instead   the generating matrix $\underline{\Sigma}: \mathbb{R}^4_p\to{\frak M}(6)$ defined as
\begin{equation}\label{SSS}
	\underline{\Sigma}(p)=\Sigma_{\frac{3}{2}}(p)+\bar pp\, \Sigma_{\frac{1}{2}}(p) \,.
\end{equation}
This matrix that reads,
	\begin{eqnarray}
		&&		\underline{\Sigma}(p^0,{\bf p})=\underline{\bar\Sigma}(p^0,-{\bf p})\nonumber\\
		&&=  {\footnotesize 
		\left[ \begin {array}{cccccc} {{ p^0_-}}^{3}&-{ p_-}{{ p^0_-}}^{2
		}&-{{ p^0_-}}^{2}\sqrt {2}{ p_-}&{ p^0_-}\sqrt {2}{{ p_-}}^{2}&{
			p^0_-}{{ p_-}}^{2}&-{{ p_-}}^{3}\\ \noalign{\medskip}-{ p_+}
		{{ p^0_-}}^{2}&{ p^0_+}{{ p^0_-}}^{2}&{ p^0_-}\sqrt {2}{ p_-}
		{ p_+}&-{ p_-}\sqrt {2}{ p^0_+}{ p^0_-}&-{{ p_-}}^{2}{ 
			p_+}&{ p^0_+}{{ p_-}}^{2}\\ \noalign{\medskip}-{{ p^0_-}}^{2}{ 
			p_+}\sqrt {2}&{ p^0_-}\sqrt {2}{ p_-}{ p_+}&{ p^0_+}{{ 
				p^0_-}}^{2}+{ p^0_-}{ p_-}{ p_+}&-{ p^0_-}{ p^0_+}{ p_-}-{{
				p_-}}^{2}{ p_+}&-{ p_-}\sqrt {2}{ p^0_+}{ p^0_-}&{ p^0_+}
		\sqrt {2}{{ p_-}}^{2}\\ \noalign{\medskip}{ p^0_-}{{ p_+}}^{2}
		\sqrt {2}&-{ p_+}\sqrt {2}{ p^0_+}{ p^0_-}&-{ p^0_-}{ p^0_+}
		{ p_+}-{ p_-}{{ p_+}}^{2}&{{ p^0_+}}^{2}{ p^0_-}+{ p_-}{
			p^0_+}{ p_+}&{ p^0_+}\sqrt {2}{ p_-}{ p_+}&-{{ p^0_+}}^{2
		}\sqrt {2}{ p_-}\\ \noalign{\medskip}{ p^0_-}{{ p_+}}^{2}&-{ 
			p_-}{{ p_+}}^{2}&-{ p_+}\sqrt {2}{ p^0_+}{ p^0_-}&{ p^0_+}
		\sqrt {2}{ p_-}{ p_+}&{{ p^0_+}}^{2}{ p^0_-}&-{{ p^0_+}}^{2}{
			p_-}\\ \noalign{\medskip}-{{ p_+}}^{3}&{ p^0_+}{{ p_+}}^{2}&{
			p^0_+}{{ p_+}}^{2}\sqrt {2}&-{{ p^0_+}}^{2}{ p_+}\sqrt {2}&-{
			{ p^0_+}}^{2}{ p_+}&{{ p^0_+}}^{3}\end {array} \right] }\,, \label{T2}
	\end{eqnarray}
 may be used for investigating some general features as it is Hermitian, $\underline{\Sigma}^+(p)=\underline{\Sigma}(p)$,  and has the remarkable property
\begin{equation}\label{masscond}
	\underline{\bar\Sigma}(p)\underline{\Sigma}(p)=\underline{\Sigma}(p)\underline{\bar\Sigma}(p)=(\bar{p}p)^3{\bf 1}\,.
\end{equation} 
Consequently, the matrices $\Sigma_s$ are Hermitian while $\Gamma_s$ are selfadjoint,
\begin{equation}
	\Sigma_s^+(p)=\Sigma_s(p)~~~\Rightarrow~~ \overline{\Gamma}_s(p)=\hat\gamma\,\Gamma_s^+(p)\hat\gamma=\Gamma_s(p)\,.
\end{equation} 
Moreover, projecting Eq. (\ref{SSS}) on  both spin subspaces we obtain the spin dependent mass conditions 
\begin{eqnarray}
	\left(\Gamma_s(p)+m^{2s}P_s \right)\left(\Gamma_s(p)-m^{2s}P_s \right)
=\left[ (\bar{p}p)^{2s}-m^{4s} \right] P_s=0\,,\label{massc}
\end{eqnarray}
giving real-valued solutions, $\bar{p}p=m^2$, only on mass-shell.

\section{Joos-Weinberg covariant field with spin $\frac{3}{2}$}

We now look for the  opportunity of  extracting the $JW(\frac{3}{2})$ field from the above approach  in which the spin sectors are mixed each other in spite of the fact that all the matrices are  maximally reducible. . 

\subsection{Separating spin sectors} 

The principal inconvenient of the   above framework  is  the interference of  spin sectors.   For example, in Eq. (\ref{Sighalf}) we see that the non-vanishing elements of the matrix $\Sigma_{\frac{1}{2}}$ span a  ${\frak M}(4)$-block   instead of a ${\frak M}(2)$ one as in the right-handed sector of Dirac's theory. This is because in the direct-product basis (\ref{basis1}) used until now the components with $\sigma=\pm\frac{1}{2}$ of the $SU(2)$ representations $(\frac{1}{2})$ and $(\frac{3}{2})$ are mixed among themselves hiding thus the physical meaning. The solution is to separate the different spin sectors in canonical  bases  (\ref{basis2a}). and (\ref{basis2b}). 

The chiral components $\phi_L$ and $\phi_R$ as well  the matrices $M\in {\frak M}(6)$ used so far are written in  direct product basis  (\ref{basis1}). We now change this basis to the canonical  bases $\{ |s,\sigma\rangle \}$ of the spaces of  representations $(\frac{1}{2})$ and $(\frac{3}{2})$ of the $SU(2)$ group. In this new basis the  fields  have new components 
\begin{equation}
\phi_L=\sum_{s,\sigma} |s,\sigma\rangle  \phi_{L\,s,\sigma}\,,\qquad   \phi_{L\,s,\sigma}=\sum_i\langle s,\sigma|\epsilon_i\rangle\langle\epsilon_i|\phi_L\rangle
\end{equation}
which in our matrix notations read $\phi_{L\,s,\sigma}=\xi^+_{s,\sigma}\phi_L$ where the $\xi$-spinors are given in Eqs. (\ref{basis2a}) and  (\ref{basis2b}). Similarly, bearing in mind that the matrices of  ${\frak M}(6)$, considered until now, have matrix elements  $M_{ij}=\langle\epsilon_i| M |\epsilon_j\rangle$, we may write the matrix elements  in canonical  basis,   $M_{s\sigma,s'\sigma'}=\xi^+_{s,\sigma}M\xi_{s',\sigma'}$. Following this procedure we find that the principal matrices of our approach are maximally reducible having matrix elements of the form     $M_{s\sigma,s'\sigma'}=\delta_{s,s'}M^s_{\sigma\sigma'}$. This procedure defines  the mappings ${\frak P}_s :{\frak M}(6) \to {\frak M}(2s+1)$ which are in fact  projections  giving $M^s={\frak P}_s (M)$. When all the chiral blocks of a matrix $X\in {\frak M}(12)$ are rewritten in  canonical  basis we obtain the new  blocks $X_L^s,X_{LR}^s,....$ that form the new matrix $X^s=\tilde{\frak P}_s (X)$ given by the new mapping $\tilde{\frak P}_s ; {\frak M}(12)\to {\frak M}(4s+2) $. Consequently, any  matrix can be represented in canonical  basis as having two diagonal blocks of spin $(\frac{1}{2}) $ and $(\frac{3}{2}) $.   In this manner we may separate the spin sectors  that now can be studied separately. 

Let us consider first the sector of spin $\frac{1}{2}$ whose principal  matrix  (\ref{Sighalf}) gives the new matrix 
	\begin{equation}
		{\frak P}_{\frac{1}{2}}(\Sigma_{\frac{1}{2}}(p))=\left(\begin{array}{cc}
			E_p-p^3&-p^1+ip^2\\
			-p^1-ip^2&E_p+p^3
			\end{array}\right)=1_{2\times2}E_p-\sigma^i p^i\,,
	\end{equation}
which is, precisely, the matrix of  left-handed sector  $(\frac{1}{2},0)$ of  Dirac's representation $\rho_D=(\frac{1}{2},0) \oplus(0,\frac{1}{2})$. More specific, in canonical  basis,  the projection matrix $\pi_{\frac{1}{2}}$ becomes the unit matrix $1_{2\times 2}$, while our $\hat\sigma$-matrices become  $\hat\sigma^0\Rightarrow 1_{2\times2}$  and  standard Pauli's ones, $\hat\sigma^i \Rightarrow\sigma^i$. Consequently, $\tilde{\frak P}_{\frac{1}{2}}(\hat\gamma^{\mu})=\gamma^{\mu}$ are Dirac's matrices in chiral  representation while $\tilde{\frak P}_{\frac{1}{2}}(\hat\gamma)= \gamma^0$ such that the adjoint defined here in the sector of spin half is just the well-known Dirac's adjoint. This means that our spin half sector coincides to Dirac's approach.  We may conclude that in  canonical  basis we have the decomposition $\hat\rho=\rho_D\oplus \rho$ where $\rho= (\frac{3}{2},0)\oplus(0,\frac{3}{2})$ is the representation of  genuine $JW(\frac{3}{2})$ field. As the spin half sector  brings nothing new we focus in what follows only on the  last one.

\subsection{Matrices in canonical  basis}

After separating the spin-half sector we remain with the $JW(\frac{3}{2})$ field   with eight degrees of freedom denoted as $\Psi$  (instead of $\psi$) for avoiding confusions. For the same reason we denote by $\rho =\rho_L\oplus\rho_R \subset {\frak M}(8)$ the representations governing their transformations. The representation spaces  of the representations $\rho_L,\, \rho_R \subset{\frak M}(4)$ are now four-dimensional. It is convenient to consider  the same basis in both these subspaces, namely  the canonical  basis (\ref{basis2a}) but in a new simpler realization, 
\begin{eqnarray}
	|\textstyle{\frac{3}{2},\frac{3}{2}}\rangle&=&(1,0,0,0)^T=\xi_{\textstyle{\frac{3}{2}}}\,, \nonumber\\
	|\textstyle{\frac{3}{2},\frac{1}{2}}\rangle&=&(0,1,0,0)^T=\xi_{\textstyle{\frac{1}{2}}}\,, \nonumber\\
	|\textstyle{\frac{3}{2},-\frac{1}{2}}\rangle&=&(0,0,1,0)^T=\xi_{\textstyle{-\frac{1}{2}}}\,, \nonumber\\
	|\textstyle{\frac{3}{2},-\frac{3}{2}}\rangle&=&(0,0,0,1)^T=\xi_{\textstyle{-\frac{3}{2}}}\,, 	\label{xi3}\label{basiscan}
\end{eqnarray}
which erase the memory of former spin half sector. These spinors are  orthonormal forming  complete systems in their representation spaces,
\begin{equation}\label{xx1}
	\xi^+_{\sigma}	\xi_{\sigma'}=\delta_{\sigma,\sigma'}\,,\qquad 
	\sum_{\sigma}	\xi_{\sigma}	\xi^+_{\sigma}={\bf 1} \,\in\, {\frak M}(4)\,.
\end{equation} 
Now we use the matrix ${\bf 1} =1_{4\times 4}$ while the matrices (\ref{gam}) and (\ref{gam5}) keep their form, 
\begin{equation}\label{gg5}
		1_{\rho}=\left(\begin{array}{cc}
		{\bf 1}&0\\
		0&{\bf 1}
	\end{array}\right)\,, \quad 
	\gamma=\left(\begin{array}{cc}
		0&{\bf 1}\\
		{\bf 1}&0
		\end{array}\right)\,, \quad 
	\gamma^5=\left(\begin{array}{cc}
		-{\bf 1}&0\\
		0&{\bf 1}
	\end{array}\right)\,,	
\end{equation}
but now in ${\frak M}(8)$.

The   representation 
 $\rho=\rho_L\oplus\rho_R=(\frac{3}{2},0)\oplus(0,\frac{3}{2})$ is formed by  single spin ones  having the generators
\begin{eqnarray}
\textstyle{(\frac{3}{2},0)} :&\qquad&\rho_L(s_i)=I_i\,,\quad \rho_L(k_i)=-iI_i\,,\\
\textstyle{(0,\frac{3}{2})} :&\qquad&\rho_R(s_i)=I_i\,,\quad \rho_R(k_i)=iI_i\,,	
\end{eqnarray}
where $I_i\equiv I_i^{\frac{3}{2}}\in{\frak M}(4)$ are the generators of the  representation $(\frac{3}{2})$ of the $SU(2)$ group, whose matrices in  canonical  basis are well-known and can be seen in Appendix B. The generators of     $\rho\subset{\frak M}(8)$ take the form
\begin{equation}
	\rho(s_j)=\left(\begin{array}{cc}
		I_j&0\\
		0&I_j
		\end{array}\right)\,,\quad \rho(k_j)=\left(\begin{array}{cc}
		-iI_j&0\\
		0&iI_j
		\end{array}\right)=i\gamma^5\rho(s_j)\,. 
\end{equation}
We now can derive the definitive form of the boosts  $\rho(l_{\bf p})=\exp[-\kappa^i\rho(k_i)]$ depending on  parameters (\ref{kapa}), 
\begin{eqnarray}
	&&	\rho_L(l_{\bf p})  =\rho_R(l^{-1}_{\bf p})=\frac{1}{(2m(E_p+m))^{\frac{3}{2}}}	\nonumber\\
&&\times	\left[ \begin {array}{cccc}  \left( { E_-}+m \right) ^{3}&- \left( 
	{ E_-}+m \right) ^{2}{ p_-}\,\sqrt {3}&{{ p_-}}^{2} \left( { 
		E_-}+m \right) \sqrt {3}&-{{ p_-}}^{3}\\ \noalign{\medskip}-\sqrt {3}
	\left( { E_-}+m \right) ^{2}{ p_+}& \frac{1}{3}\left( { E_-}+m
	\right) A&-\frac{1}{3}\,{ p_-}\,B&\sqrt {3}{{ p_-}}^{2} \left( { E_+}+m
	\right) \\ \noalign{\medskip}\sqrt {3} \left( { E_-}+m \right) {{
			p_+}}^{2}&-\frac{1}{3}\,{ p_+}\,B& \frac{1}{3}\left( { E_+}+m \right) A&-
	\sqrt {3} \left( { E_+}+m \right) ^{2}{ p_-}\\ \noalign{\medskip}-
	{{ p_+}}^{3}&{{ p_+}}^{2} \left( { E_+}+m \right) \sqrt {3}&-
	\left( { E_+}+m \right) ^{2}{ p_+}\,\sqrt {3}& \left( { E_+}+m
	\right) ^{3}\end {array} \right]\,, \label{T3}
\end{eqnarray}
where
\begin{equation} \label{T3a}
	\begin{array}{l}
		A=9{ E_+}{ E_-}+3m({ E_-}+{ E_+})-3{m}^{2}\,,\\
		B=9{ E_+}\,{ E_-}+6m({ E_-}+{ E_+})+3{m}^{2}\,,
	\end{array}	
\end{equation}  
 tanks to our  mappings ${\frak P}_s$  allowing us to express  $\rho_{L/R}(l_{\bf p})= {\frak P}_{\frac{3}{2}} \left(\hat\rho_{L/R}(l_{\bf p})\right)$ 
in terms of  matrices (\ref{T1})   derived previously as  in Appendix A. Therefore, we may conclude that our method helps us to compute the sums the exponential series $\rho_{L/R}(l_{\bf p})=\exp(\mp\kappa^iI_i)$. We may convince ourselves that this is true expanding the matrices (\ref{T3}) as
\begin{eqnarray}
	\rho_{L/R}(l_{\bf p})={\bf 1}&+& p^i\left[\partial_{p^i}\rho_{L/R}(l_{\bf p})\right]_{{\bf p}=0}+	 \frac{1}{2}	p^i p^j\left[\partial_{p^i}\partial_{p^j}\rho_{L/R}(l_{\bf p})\right]_{{\bf p}=0}+\cdots \nonumber\\
	&=&{\bf 1}\mp\frac{1}{m} p^i I_i+\frac{1}{2 m^2}(p^i I_i)^2 +\cdots\,,
\end{eqnarray}
bearing in mind that the parameters (\ref{kapa})  behave as $\kappa^i=\frac{p^i}{m}+O(|{\bf p}|^3)$.

Of a special interest is the  matrix 
	\begin{eqnarray}
		\Sigma(p^0,{\bf p})=\bar\Sigma (p^0,-{\bf p})=	\left[ \begin {array}{cccc} {{ p^0_-}}^{3}&-{ p_-}{{ p^0_-}}^{2}
		\sqrt {3}&{ p^0_-}{{ p_-}}^{2}\sqrt {3}&-{{ p_-}}^{3}
		\\ \noalign{\medskip}-\sqrt {3}{ p_+}{{ p^0_-}}^{2}&{ p^0_-}
		\left( { p^0_+}{ p^0_-}+2{ p_-}{ p_+} \right) &-{ p_-}
		\left( 2{ p^0_+}{ p^0_-}+{ p_-}{ p_+} \right) &\sqrt {3}{
			p^0_+}{{ p_-}}^{2}\\ \noalign{\medskip}\sqrt {3}{ p^0_-}{{ 
				p_+}}^{2}&-{ p_+} \left( 2{ p^0_+}{ p^0_-}+{ p_-}{ p_+}
		\right) &{ p^0_+} \left( { p^0_+}{ p^0_-}+2{ p_-}{ p_+}
		\right) &-\sqrt {3}{{ p^0_+}}^{2}{ p_-}\\ \noalign{\medskip}-{{ 
				p_+}}^{3}&{ p^0_+}{{ p_+}}^{2}\sqrt {3}&-{{ p^0_+}}^{2}{ p_+}
		\sqrt {3}&{{ p^0_+}}^{3}\end {array} \right] \,,\label{T4}
	\end{eqnarray}
defined as  the projection $\Sigma ={\frak P}_{\frac{3}{2}}(\underline{\Sigma})$  of  the generating matrix (\ref{SSS}).   This matrix is Hermitian and 
has similar properties as the generating matrix determining the properties of the matrix
\begin{equation}\label{ggg1}
\Gamma(p)=\Gamma(p^0,{\bf p})=\left(\begin{array}{cc}
	0&\Sigma(p_0,{\bf p})\\
	\Sigma(p_0,-{\bf p})&0
	\end{array}    \right)	\in {\frak M}(8)
\end{equation}
off mass-shell. Thus we  may verify  the  identities
\begin{equation}
	\Sigma(p_0,{\bf p})\Sigma(p_0,-{\bf p}) =(\bar p p)^3{\bf 1}~~ \Rightarrow~~
	\Gamma(p)^2=(\bar p p)^3 1_{\rho}\,,
\end{equation}
that help us to rewrite the mass condition  (\ref{massc}) as
\begin{equation}\label{ultimass}
(\Gamma(p)+m^31_{\rho})(\Gamma(p)-m^31_{\rho})=[(\bar p p)^3-m^6]1_{\rho}=0\,,
\end{equation}
or even to derive Fourier transforms of  Green functions.

On mass-shell we verify the identity
\begin{equation}
\Sigma(E_p,{\bf p})\rho_R(l_{\bf p})	=m^3\rho_L(l_{\bf p})
\end{equation}
which solves the field equation in momentum representation,  
\begin{equation}\label{EQQ}
	\Gamma(E_p,{\bf p})\rho(l_{\bf p})=m^3\rho(l_{\bf p})\,,
\end{equation}
up to the spinors in rest frame.

On the other hand, working off mass-shell we may use the matrices $\Sigma(p)=\Sigma(p^0,{\bf p})$ and $\bar\Sigma(p)=\Sigma(p^0,-{\bf p})$  for deriving  new $\sigma$-matrices using a similar definition to (\ref{sig1})  as
\begin{eqnarray}
	\sigma^{\alpha\beta\gamma}=\bar\sigma_{\alpha\beta\gamma}=\frac{1}{3!}\partial_{p_{\alpha}}\partial_{p_{\beta}}\partial_{p_{\gamma}}\Sigma(p)=\frac{1}{3!}\partial_{p^{\alpha}}\partial_{p^{\beta}}\partial_{p^{\gamma}}\bar\Sigma(p)\,.
\end{eqnarray} 
These matrices which are now from ${\frak M}(4)$ will be denoted without hat  for distinguish them from the former $\hat\sigma$- matrices of ${\frak M}(6)$. We obtain thus a set of 20 Hermitian matrices, symmetric in all their indices. However,  observing that $\eta_{\mu\nu}\sigma^{\mu\nu\sigma}=0$ we conclude that we remain with 16 linear independent matrices for  building  $\gamma$-matrices which will satisfy a similar condition, $\eta_{\mu\nu}\gamma^{\mu\nu\sigma}=0$. This could explain why we do not have  a first order equation here.

We constructed our approach starting with $\sigma^{000}={\bf 1}$. The other matrices  with equal indices, presented in Appendix B,  are square roots, $(\sigma^{111})^2= (\sigma^{222})^2=(\sigma^{333})^2={\bf 1}$. 
These matrices are unimodular but the other ones  have  the same determinant but of another value, 
\begin{equation}
	\det(\sigma^{\alpha\beta\gamma})=\det(\sigma_{\alpha\beta\gamma})=\left\{\begin{array}{ll}
			1\quad&\alpha=\beta=\gamma\\
			\frac{1}{9}& {\rm otherwise}
			\end{array}
			\right. \,.
\end{equation}
The matrices $\sigma^{00j}$ and $\sigma^{0ij}$ have special properties, 
\begin{equation}\label{ssi}
\sigma^{00j}=-\sigma_{00j}=\frac{2}{3}I_j\,,~~ 
\sigma^{0ij}=\sigma_{0ij}=\frac{1}{3}\left\{I_i, I_j\right\}-\frac{1}{2}\delta_{ij}{\bf 1}\,,
\end{equation}   
relating the algebra of $\sigma$-matrices to the Lie algebra of the irreducible representation $(\frac{3}{2})$ of the $SU(2)$ group. The algebraic properties of $\sigma$-matrices determine those of  the matrices 
\begin{equation}\label{ggg}
	\gamma^{\alpha\beta\gamma}=\left(\begin{array}{cc}
		0&\sigma^{\alpha\beta\gamma}\\
		\bar\sigma^{\alpha\beta\gamma}&0
	\end{array}\right)	\in {\frak M}(8) \,,
\end{equation}
that are the matrices (\ref{maga}) rewritten in canonical  basis (and denoted without hat). These have interesting properties as, for example, in the case of the matrices
\begin{equation}
	\gamma^{00j}=\left(\begin{array}{cc}
	0&\sigma^{00j}\\
	-\sigma^{00j}&0
	\end{array}\right)\,,\quad \gamma^{0ij}=\left(\begin{array}{cc}
		0&\sigma^{0ij}\\
		\sigma^{0ij}&0
	\end{array}\right)\,, 
\end{equation}	
which comply with the identities   
\begin{eqnarray}
&&		\left[  \gamma^{00i}, \gamma^{00j}  \right]=-i\frac{4}{9}\,\varepsilon_{ijk}\rho(s_k) \,,\\
&&	\left\{  \gamma^{00i}, \gamma^{00j}\right\} +\frac{4}{3}\gamma^{000}\gamma^{0ij}=\frac{2}{3}\delta_{ij}1_{\rho} \,.
\end{eqnarray}
More refined properties could be found after an extended  study that could reveal new interesting mathematical properties.

\subsection{Normalized mode spinors}

Turning back to configuration representation we write the  third order equation of the $JW(\frac{3}{2})$ field  as,
\begin{equation}\label{EEE}
	(i)^3\gamma^{\mu\nu\sigma}\partial_{\mu}\partial_{\nu}\partial_{\sigma} \Psi-m^3\Psi =0\,,
\end{equation}
using the new $\gamma$-matrices (\ref{ggg}). Looking for  general solutions
of the form
\begin{eqnarray}\label{Psi1}
		\Psi _{\varepsilon} (x)=	\Psi _{\varepsilon}^+(x)+	\Psi_{\varepsilon}   ^-(x)=\int_{p\in\Omega_{\mathring p}} d^3p \sum_{\sigma}\left[U^{\varepsilon} _{{\bf  p},\sigma}(x) \alpha_{\sigma}({\bf  p}) +V^{\varepsilon} _{{\bf  p}, \sigma}(x) \beta^ {*} _{\sigma}({\bf  p})\right]\,,
\end{eqnarray}
expressed in terms of wave spinors,  $\alpha $ and   $\beta $, and   8-component mode spinors, 
\begin{eqnarray}
	U^{\varepsilon}_{{\bf  p},\sigma}(x)&=&u_{\sigma}({\bf  p})\frac{1}{(2\pi)^{\frac{3}{2}}} \,e^{-i\varepsilon E_pt+i\varepsilon{\bf  p}\cdot{\bf  x}}\,,\label{U1} \\
	V^{\varepsilon}_{{\bf  p},\sigma}(x)&=&v_{\sigma}({\bf  p})\frac{1}{(2\pi)^{\frac{3}{2}}}\, e^{i\varepsilon E_pt-i\varepsilon{\bf  p}\cdot{\bf  x}}\,,\label{V1}
\end{eqnarray}
similar to the 12-dimensional ones,  (\ref{U}) and (\ref{V}), but  introducing the new parameter $\varepsilon$ to distinguish among the different solutions of the mass condition (\ref{ultimass}). This parameter is a cubic root of unity  denoted  as
\begin{equation}
	\varepsilon^3=1~~\Rightarrow~~ \varepsilon_0=1\,,\quad \varepsilon_{\pm}=-\frac{1}{2}\pm i\frac{\sqrt{3}}{2}\,,
\end{equation}
which do not affect the equations in momentum representation that read
\begin{eqnarray}
	\Gamma(p) u_{\sigma}({\bf p})-m^{3}u_{\sigma}({\bf p})&=&0\,,	~\\
	~~\Gamma(p) v_{\sigma}({\bf p})+m^{3}v_{\sigma}({\bf p})&=&0\,,
\end{eqnarray}
where $\Gamma(p)=\gamma^{\mu\nu\sigma}p_{\mu}p_{\nu}p_{\sigma}~\in{\frak M}(8)$ is the matrix (\ref{ggg1}) on mass-shell, with $p\in \Omega_{\mathring{p}}$. Note that the spinors $u_{\sigma}$ and $v_{\sigma}$ do not depend on $\varepsilon$ as these are in fact functions of $\frac{p}{m}$. We conclude  that the spinors  (\ref{U1}) and  (\ref{V1}) are solutions of  the third order  equations ({\ref{EEE}}) but satisfying specific mass conditions,
\begin{equation}
 \left[ \left(	\partial_{\mu}\partial^{\mu}\right)^3+m^6\right]  U^{\varepsilon}_{{\bf  p},\sigma}(x)=0 \quad    \Rightarrow\quad  \left(	\partial_{\mu}\partial^{\mu}+\varepsilon^2m^2\right)U^{\varepsilon}_{{\bf  p},\sigma}(x)=0\,,\quad \varepsilon=1,\,\varepsilon_{+},\,\varepsilon_-\,,
\end{equation} 
and similarly for the spinors $V^{\varepsilon}_{{\bf  p},\sigma}(x)$.
 
 On the other hand, the identity (\ref{EQQ}) guarantees that the solutions of the field equation  in momentum representation have the general form  $\rho(l_{\bf p}) \zeta$ where $\zeta$ may be any spinor dependent or independent on $p$. As in rest frames where ${\bf p}=0$ the spinors $u_{\sigma}(0)$ and $v_{\sigma}(0)$ are eigenvectors of the matrix $\Gamma(\mathring{p})=\gamma^{000}=\gamma$ given in Eq. (\ref{gg5}), $\gamma u_{\sigma}(0)=u_{\sigma}(0)$ and $\gamma v_{\sigma}(0)=-v_{\sigma}(0)$, we may define 
\begin{eqnarray}
	{u}_{\sigma}(0) =\frac{1}{\sqrt{2}} \left(\begin{array}{c}
		\xi_{\sigma}\\
		\xi_{\sigma}
	\end{array}\right)\,,\quad
	{v}_{\sigma}(0)=\frac{1}{\sqrt{2}}\left(
	\begin{array}{c}
		\eta_{\sigma}\\
		-\eta_{\sigma}
	\end{array}\right)\,,\label{xy1}
\end{eqnarray}
where $\xi_{\sigma}$ are the spinors (\ref{basiscan}) of  canonical  basis  while $\eta_{\sigma}=(-1)^{s+\sigma}\xi_{ -\sigma}$ are the spinors of  equivalent complex conjugated irreducible representations, $(s)^*$ of the antiparticle sector \cite{JW2,W}. We obtain thus the form of  spinors in momentum representation,
\begin{eqnarray}
	u_{\sigma}({\bf  p})&=&n({p})\rho(l_{\bf  p})u_{\sigma}(0)=\frac{n(p)}{\sqrt{2}} \left(\begin{array}{c}
		\rho_L(l_{\bf  p})	\xi_{\sigma}\\
		\rho_R(l_{\bf  p})	\xi_{\sigma}
	\end{array}\right)  \,,  \label{Ufin} \\  
	v_{\sigma}({\bf  p})&=&n(p)\rho(l_{\bf  p}){v}_{\sigma}(0)=\frac{n(p)}{\sqrt{2}} \left(\begin{array}{c}
		\rho_L(l_{\bf  p})	\eta_{ \sigma}\\
		-\rho_R(l_{\bf  p})\eta_{ \sigma}
	\end{array}\right)  \,,    \label{Vfin}
\end{eqnarray}
depending on the matrices (\ref{T3}) and a normalization factor  which must satisfy the condition $n(\mathring{p})=1$ in rest frames.

We now may  test how our new inner product (\ref{sp})  works when we try to normalize the mode spinors. Let us consider first  the mode spinors with $\varepsilon=1$ we denote simply,  $U^{\varepsilon=1}_{{\bf  p},\sigma}(x)=U_{{\bf  p},\sigma}(x)$ and $V^{\varepsilon=1}_{{\bf  p},\sigma}(x)=V_{{\bf  p},\sigma}(x)$.  The inner product 
\begin{equation}
	\langle U_{{\bf  p},\sigma},U_{{\bf  p}',\sigma'}\rangle =\delta({\bf p}-{\bf p}')
	\overline{u}_{\sigma}({\bf p}) Q (p,p') u_{\sigma'}(\bf p')
\end{equation}
depends on the matrix
\begin{equation}
	Q(p,p')=\frac{1}{3 m^2}\left( p_{\mu}p'_{\nu} +p_{\mu}p_{\nu}+p'_{\mu}p'_{\nu}     \right) \gamma^{0\mu\nu}\,,
\end{equation}
which results after performing the derivatives of Eq. (\ref{sp}). As in this case we have  $p=p' \in\Omega_{\mathring{p}}$ we may write
\begin{equation}
	Q(p,p)=\frac{1}{m^2}\gamma^{0\mu\nu}p_{\mu}p_{\nu}\,,
\end{equation}
observing that  the matrix we need can be derived easily as
\begin{equation}
	\gamma^{0\mu\nu}p_{\mu}p_{\nu}=\left.\frac{1}{3}\partial_{p^0}\Gamma(p^0,{\bf p})\right|_{p^0=E_p}=\left(\begin{array}{cc}
		0&\theta(p)\\
		\bar\theta(p)&0
		\end{array}\right)\,,
\end{equation}
where 
\begin{eqnarray}\label{theta}
&&	\theta(E_p,{\bf p})=\bar\theta(E_p,-{\bf p})=\left.\frac{1}{3}\partial_{p^0}\Sigma(p^0,{\bf p})\right|_{p^0=E_p}
\nonumber\\
	&&=\left[ \begin {array}{cccc} {{ E_-}}^{2}&-\frac{2}{\sqrt{3}}{ p_-}{ E_-}
	&\frac{1}{\sqrt{3}}{{ p_-}}^{2}&0\\ \noalign{\medskip}-\frac{2}{\sqrt{3}}{ p_+}{ E_-}&~~\frac{1}{3}{{ E_-}}^{2}+\frac{4}{3}{ E_-}{ E_+
	}-\frac{2}{3}{m}^{2}&-\frac{2}{3}({ p_-}{ E_-}+{ p_-}{ E_+})&\frac{1}{\sqrt{3}}{{ p_-}}^{2}\\ \noalign{\medskip}\frac{1}{\sqrt{3}}{{ p_+}}^{
		2}&-\frac{2}{3}({ E_-}{ p_+}+{ E_+}{ p_+})&\frac{4}{3}{ E_-}{ E_+}+\frac{1}{3}{{ E_+}}^{2}-\frac{2}{3}{m}^{2}&-\frac{2}{\sqrt{3}}{ E_+}{
		p_-}\\ \noalign{\medskip}0&\frac{1}{\sqrt{3}}{{ p_+}}^{2}&-\frac{2}{\sqrt{3}}{ 
		E_+}{ p_+}&{{ E_+}}^{2}\end {array} \right]\,, \label{T5}
\end{eqnarray}
as it results from  Eq.  (\ref{ggg1}).  Gathering now all the above results we obtain
\begin{eqnarray}
	\langle U_{{\bf  p},\sigma},U_{{\bf  p}',\sigma'}\rangle =\delta^3({\bf p}-{\bf p}')\frac{n(p)^2}{2m^2}	\xi^+_{\sigma}\left[\rho_R(l_{\bf p}) \theta(p)\rho_R(l_{\bf p})    +\rho_L(l_{\bf p})\bar \theta(p)\rho_L(l_{\bf p})\right]\xi_{\sigma'}\,.
\end{eqnarray}
The  matrix $[\rho_R(l_{\bf p}) \theta(p)\rho_R(l_{\bf p})+...]$ can be calculated using Eqs.  (\ref{T3})  and (\ref{T3a}) obtaining the final result  $2m E_p{\bf 1}$. Therefore,  we must take the normalization factor $n(p)=\sqrt{\frac{m}{E_p}}$ which complies with the condition $n(\mathring{p})=1$. As  the spinors $\xi_{\sigma}$ form the orthonormal system   (\ref{xx1}) we obtain 
\begin{equation}
\langle U_{{\bf  p},\sigma},U_{{\bf  p}',\sigma'}\rangle =\delta^3({\bf p}-{\bf p}')\delta_{\sigma,\sigma'}\,.	
\end{equation}
Similar normalization can be found for the mode spinors of negative frequency $V_{{\bf  p},\sigma}$.

It remains to verify the orthogonality of  particle and antiparticle subspaces calculating 
\begin{equation}
	\langle V_{{\bf  p},\sigma},U_{{\bf  p}',\sigma'}\rangle =\delta^3({\bf p}+{\bf p}')
	\overline{v}_{\sigma}({\bf p}) Q (p,p') u_{\sigma'}(\bf p')\,,
\end{equation}
where now $p'=(E_p,-{\bf p})$.  Consequently, all the terms linear in $p^i$ of the matrix $Q(p,p')$ vanish such that $\theta(E_p,{\bf p})=\theta(E_p,-{\bf p})$ which means that  $\theta(p)=\theta_S(p)=\frac{1}{2}(\theta(p)+\bar\theta(p))$. With this notation we obtain   
\begin{eqnarray}
		\langle V_{{\bf  p},\sigma},U_{{\bf  p}',\sigma'}\rangle &=&\delta^3({\bf p}+{\bf p}')\frac{1}{2m E_p}\eta^+_{\sigma}\left[\rho_L(l_{\bf p}) \theta_S(p)\rho_R(l_{\bf p})  -\rho_R(l_{\bf p}) \theta_S(p)\rho_L(l_{\bf p})\right]\xi_{\sigma'} =0 \,,
\end{eqnarray}
verifying that the particle and antiparticle subspaces are orthogonal with respect to our inner product (\ref{sp}). The conclusion is that this inner product assures the normalization of  mode spinors,
\begin{eqnarray}
	\langle U_{{\bf p},\sigma}, U_{{{\bf p}\,}',\sigma'}\rangle &=&
	\langle V_{{\bf p},\sigma}, V_{{{\bf p}\,}',\sigma'}\rangle=	\delta_{\sigma\sigma^{\prime}}\delta^{3}({\bf p}-{\bf p}\,^{\prime})\,,\label{ortU}\\
	\langle U_{{\bf p},\sigma}, V_{{{\bf p}\,}',\sigma'}\rangle &=&
	\langle V_{{\bf p},\sigma}, U_{{{\bf p}\,}',\sigma'}\rangle =0\,, \label{ortV}
\end{eqnarray}
in spite of its unusual form. 

When we know the physical  field $\Psi$ we may derive the wave spinors $\alpha\in \tilde{\cal F}^{+}$ and $\beta\in \tilde {\cal F}^{-}$ applying the inversion formulas
\begin{equation}\label{inv}
	\alpha_{\sigma}({\bf p})=\langle U_{{\bf p},\sigma},\Psi \rangle\,, \qquad 
	\beta_{\sigma}({\bf p})=\langle \Psi,  V_{{\bf p},\sigma} \rangle \,,
\end{equation}
resulted from  Eqs. (\ref{ortU}) and (\ref{ortV}). In general, the spaces $\tilde{\cal F}^+\sim \tilde{\cal F}^-$ are rigged Hilbert spaces, including Hilbert spaces ${\cal L}^2(\Omega_{\mathring{p}}, d^3p,{\cal V}_{\frac{3}{2}})$,  equipped with the same scalar product, 
\begin{eqnarray}\label{spa}
	\langle \alpha, \alpha'\rangle=\int d^3p \sum_{\sigma}\alpha_{\sigma}^*({\bf p})	\alpha_{\sigma}'({\bf p})\,,
\end{eqnarray}
and similarly for the spinors $\beta$. Then after using  Eqs.  (\ref{ortU}) and (\ref{ortV}) we obtain  the important identity 
\begin{equation}\label{spp}
	\langle \Psi  , \Psi  '\rangle =\langle \alpha,\alpha'\rangle + \langle \beta,\beta'\rangle	\,,
\end{equation}
expressing the  inner product in terms of wave spinors.  We have thus all we need for verifying that the wave spinors transform under Wigner's induced representation  \cite{Wig} of spin $\frac{3}{2}$ which is unitary with respect to the scalar product (\ref{spa}) \cite{Mc}.

However, these properties hold only for the fields with $\varepsilon=1$ which are suitable for quantization. Otherwise,  in the cases of  the fields $\Psi_{\varepsilon_{\pm}}$ we observe that  all the space integrals arising in inner products are divergent such that these fields cannot get  physical meaning having infinite norms.   Nevertheless, these  solutions remain of mathematical interest in analyzing the properties of  field equation off mass-shell. For example,  the form of propagator is determined now by six poles in the complex plane $p^0$, namely $\pm E_p$, $\pm  \sqrt{{\bf p}^2+\varepsilon_+ m^2}$ and  $\pm  \sqrt{{\bf p}^2+\varepsilon_- m^2}$.

\section{Concluding remarks}

We succeeded to develop the  explicit theory of  the $JW(\frac{3}{2})$ field  tanks to our new method based on maximally reducible representations allowing us to write down the matrices of the Wigner boosts in canonical basis. These fields satisfy equations of third order coming from a more complicated Lagrangian formalism where we defined the new inner product (\ref{sp}) which behaves remarkably in normalizing the eight-dimensional mode spinors with physical meaning. Another specific feature of this approach is the mass condition  (\ref{ultimass}) which has only one real valued solution, $\bar p p-m^2=0$, such that it seems there are no dipole or multi-pole ghosts states as in  theories leading to mass conditions  of the form $\left( \bar p p-m^2\right)^n=0$ \cite{G1,G2}. However,  instead of them, we encounter nonphysical solutions of infinite norm, $\Psi_{\varepsilon_{\pm}}$,   that may be ignored at least at the level of free theory but it is premature to discuss now what happens when interactions are introduced.

 For testing how the physical field $\Psi$ interacts with other fields we must consider  suitable couplings leading to a renormalizable theory of interacting fields. In the case of our approach based on a third order equation this condition is non trivial as, for example,  the minimal coupling with the electromagnetic field may generate non-renormalizable terms.  Therefore,  we must restrict ourselves to the interaction Lagrangian density  $\propto j^{\mu}A_{\mu}$, in which the electromagnetic potential couples the current density (\ref{cur}).  As the current density  contains now  first and second order derivatives we do not exclude a new phenomenology we hope to be free of inconsistencies and anomalies. Interesting couplings  with tensor fields  are expected in Minkowski space-time but especially in general relativity where the flat condition $\eta_{\mu\nu}\gamma^{\mu\nu\sigma}=0$ may select special couplings with the gravitational field.

Finally, let us discuss how the method of maximally reducible representations  can be generalized to any set of fields with half integer spin, from $\frac{1}{2}$  up to the maximal spin   $s_{\rm m}=\frac{1}{2}+k$ where $k$ is an arbitrary integer. We define first the chiral  representations
\begin{eqnarray}
\hat\rho_L&=&(1,0)^{\otimes k}\otimes (\textstyle{\frac{1}{2}},0)=(s_{\rm m},0)\oplus (s_{\rm m}-1,0)\oplus \cdots\,, \nonumber\\
	\hat\rho_R&=&(0,1)^{\otimes k}\otimes (0,\textstyle{\frac{1}{2}})=
		(0,s_{\rm m})\oplus (0,s_{\rm m}-1)\oplus \cdots\,,\label{dec}
\end{eqnarray}
of dimension $n=2\times 3^k$, formed as Kronecker products of simple representations presented  in Appendix A.  In subspaces ${\cal V}_L$ and ${\cal V}_R$ we may introduce the common direct product basis  $\{|\epsilon_i\rangle ,  i=1,2,\cdots n\}$ with a convenient enumeration. In this basis the matrices $M\in {\frak M}(n)$ of chiral sectors have matrix elements $M_{ij}=\langle i|M|j\rangle$.  The next step is to  change this basis with the canonical basis of the representation $(s_{\rm m})\oplus\cdots (s)_1\oplus(s)_2\oplus\cdots (s)_{n_s}\oplus\cdots$, resulted after decomposing the chiral representations as in Eq. (\ref{dec}). Each representation $(s)$ is supposed to have the multiplicity $n_s$ which means that a supplemental index, $\iota_s$ , is needed for distinguish among the representations  with same spin $s$.  
Furthermore, projection operators ${\frak P}_{\iota_s,s}: {\frak M}(n)\to {\frak  M}(2s+1)$ separating spin sectors can be defined such that any matrix $M\in {\frak M}(n)$ having the  elements $M_{ij}$ be projected to the matrix $M^{\iota_s,s}={\frak P}_{\iota_s,s}(M)$ with the matrix elements
\begin{equation}
	 M^{\iota_s,s}_{\sigma,\sigma'}
	=\sum_{i,j}\langle \iota_s,s,\sigma |i \rangle\langle i|M|j\rangle\langle j|\iota_s,s,\sigma'\rangle \,.
\end{equation}
Using these projection operators, the entire framework can be split in independent  sectors governed by  representations $(s,0)\oplus(0,s)$ with $\frac{1}{2}\le s \le s_{\rm m}$,  obtain thus the corresponding set of  $JW(s)$  fields. As there are no major difficulties in applying this method we hope to obtain closed formulas of all  quantities we need for writing down explicitly the Feynman rules for any half integer spin.

\appendix

\section*{Appendx A: Building boosts}

	\renewcommand{\theequation}{A\arabic{equation}}
	\setcounter{equation}{0} \renewcommand{\theequation}
	{A.\arabic{equation}}
		
			The transformations $l_{\bf p}\in SL(2,\mathbb{C})/SU(2)$  are parametrized as
		\begin{equation}\label{kapa}
			l_{\bf p}=e^{-i\kappa^i k_i}\,, \quad\kappa^i=\frac{p^i}{|{\bf p}|}{\rm tanh}^{-1} \frac{|{\bf p}|}{E_p}\,,\quad~~  l^{-1}_{\bf  p}=l_{-{\bf p}} \,,
		\end{equation}
		giving rise to the Lorentz boosts $\rho_V(l_{\bf  p})=\Lambda(l_{\bf  p})$ having the  matrix elements,  
		\begin{eqnarray}
			\Lambda^{0\cdot}_{\cdot 0}(l_{\bf p})&=&\frac{E_p}{m},\quad 	\Lambda^{0\cdot}_{\cdot i}(l_{\bf p})=	\Lambda^{i\cdot}_{\cdot 0}(l_{\bf p})=\frac{p^i}{m},\label{Lal}\\ 
				\Lambda^{i\cdot}_{\cdot j}(l_{\bf p})&=&\delta_{ij}+\frac{p^i p^j}{m(E_p+m)},~~\label{Lboost}
		\end{eqnarray}
		which transforms the representative momentum $\mathring{p}=(m,0,0,0)$  into the desired momentum $p=(E_p,{\bf  p})=\Lambda(l_{\bf  p})\,\mathring{p}$.  
		
		The  representation $\tau_{\frac{1}{2}}=(\frac{1}{2},0)$ is the left-handed  part of Dirac's one such that we may write
		\begin{equation}
		\tau_{\frac{1}{2}}(\l_{\bf p})	=\frac{1}{\sqrt{2m(E_p+m)}}\left(\begin{array}{cc}
			E_-+m&-p_-\\
			-p_+&E_++m
			\end{array}\right)\,,
		\end{equation}
		where we used  null momentum components on mass-shell, 
		\begin{equation}\label{nul1} 
			E_{\pm}=E_p\pm p^3\,,\quad p_{\pm}=p^1\pm ip^2~~ \Rightarrow 
			~~\bar{p} p=E_+E_--p_+p_- \,.
		\end{equation}
			Note that off mass-shell we must consider $p^0_{\pm} =p^0\pm p^3$ instead of $E_{\pm}$.
			
			The boost generators of $\tau_1=(1,0)$,   $\tau_1(k_i)=-iI^1_i$,  depend on the generators $I^1_i$ of  representation $(1)$ of the $SU(2)$ group whose matrices in canonical  basis read
	\begin{equation}
		I^1_1=\frac{1}{\sqrt{2}}\left(\begin{array}{ccc}
			0&1&0\\
			1&0&1\\
			0&1&0
			\end{array}\right)\,,\quad
				I^1_2=\frac{1}{\sqrt{2}}\left(\begin{array}{ccc}
				0&-i&0\\
				i&0&-i\\
				0&i&0
			\end{array}\right)\,,\quad
				I^1_3=\left(\begin{array}{ccc}
				1&0&0\\
				0&0&0\\
				0&0&-1
			\end{array}\right)\,.
	\end{equation}
	Using  these generators we  compute	
	\begin{eqnarray}
		\tau_1(l_{\bf p})=e^{-\kappa^i I^1_i}=\frac{1}{2m(E_p+m)}\left(\begin{array}{ccc}
		(E_-+m)^2&-\sqrt{2}(E_-+m)p_-&(p_-)^2\\
		-\sqrt{2}(E_-+m)p_+&E_+E_-+mE_p&-\sqrt{2}(E_++m)p_-\\
	(p_+)^2	&-\sqrt{2}(E_++m)p_+&(E_++m)^2	
		\end{array}\right)\,,~~~~
	\end{eqnarray}
in terms of null momentum components defined above.

With these elements we construct the matrices  (\ref{T1}) defined as 
\begin{equation}\label{LRmat}
\hat\rho_L(l_{\bf p})= \hat\rho_R(l_{\bf p})^{-1}=  \tau_1(l_{\bf p})\otimes \tau_{\frac{1}{2}}(\l_{\bf p})\,,
\end{equation}
using the Kronecker product in indicated order.  Finally we verify that these matrices are Hermitian, $\hat\rho_{L/R}^+=\hat\rho_{L/R}$, and are maximally reducible, 
\begin{equation}
\hat\rho_{L/R}(l_{\bf p})=	\hat\rho_{L/R}(l_{\bf p}) \pi_{\frac{3}{2}}+
\hat\rho_{L/R}(l_{\bf p}) \pi_{\frac{1}{2}}\,,
\end{equation}	
acting separately on the subspaces of spin $\frac{3}{2}$ and $\frac{1}{2}$.

\section*{Appendix B:  $ (\frac{3}{2})$ generators and $\sigma$-matrices}

\renewcommand{\theequation}{B\arabic{equation}}
\setcounter{equation}{0} \renewcommand{\theequation}
{B.\arabic{equation}}

The generators of  irreducible representation $(\frac{3}{2})$ of the $SU(2)$ group, 
\begin{eqnarray}
I_1&=&\left[ \begin {array}{cccc} 0&\frac{\sqrt{3}}{2}&0&0\\ \noalign{\medskip}
\frac{\sqrt{3}}{2}&0&1&0\\ \noalign{\medskip}0&1&0&\frac{\sqrt{3}}{2}
\\ \noalign{\medskip}0&0&\frac{\sqrt{3}}{2}&0\end {array} \right] \,,\quad
I_2=\   \left[ \begin {array}{cccc} 0&-i\frac{\sqrt{3}}{2}&0&0\\ \noalign{\medskip}i\frac{\sqrt{3}}{2}&0&-i&0\\ \noalign{\medskip}0&i&0&-i\frac{\sqrt{3}}{2}
\\ \noalign{\medskip}0&0&i\frac{\sqrt{3}}{2}&0\end {array} \right] \,,\\
I_3&=& {\rm diag}\left(\textstyle{\frac{3}{2},\frac{1}{2},-\frac{1}{2},-\frac{3}{2}}\right)\,,
\end{eqnarray}
are related to the $\sigma$-matrices (\ref{ssi}).  Other interesting matrices 
\begin{eqnarray}
\sigma^{111}&=& \left[ \begin {array}{cccc} 0&0&0&~1\\ \noalign{\medskip}0&0&~1&0
\\ \noalign{\medskip}0&~1&0&0\\ \noalign{\medskip}~1&0&0&0\end {array}
\right]     \,,\quad
\sigma^{222}=  \left[ \begin {array}{cccc} 0&0&0&i\\ \noalign{\medskip}0&0&-i&0
\\ \noalign{\medskip}0&i&0&0\\ \noalign{\medskip}-i&0&0&0\end {array}
\right]     \,,\\
\sigma^{333}&=&{\rm diag}(1,-1,1,-1)\,,
\end{eqnarray}
\begin{eqnarray}
\sigma^{112}&=&\left[ \begin {array}{cccc} 0&0&0&-i\\ \noalign{\medskip}0&0&-i\frac{1}{3}&0
\\ \noalign{\medskip}0&i\frac{1}{3}&0&0\\ \noalign{\medskip}i&0&0&0\end {array}
\right]     \,, \quad
\sigma^{113} =\left[ \begin {array}{cccc} 0&0&\frac{1}{\sqrt{3}}&0\\ \noalign{\medskip}0
&\frac{2}{3}&0&-\frac{1}{\sqrt{3}}\\ \noalign{\medskip}\frac{1}{\sqrt{3}}&0&-\frac{2}{3}&0
\\ \noalign{\medskip}0&-\frac{1}{\sqrt{3}}&0&0\end {array} \right]     
\end{eqnarray}
\begin{eqnarray}
\sigma^{221}&=&\left[ \begin {array}{cccc} 0&0&0&-1\\ \noalign{\medskip}0&0&\frac{1}{3}&0
\\ \noalign{\medskip}0&\frac{1}{3}&0&0\\ \noalign{\medskip}-1&0&0&0
\end {array} \right]  \,, \quad 
\sigma^{223}= \left[ \begin {array}{cccc} 0&0&-\frac{1}{\sqrt{3}}&0
\\ \noalign{\medskip}0&\frac{2}{3}&0&\frac{1}{\sqrt{3}}\\ \noalign{\medskip}-\frac{1}{\sqrt{3}}&0&-\frac{2}{3}&0\\ \noalign{\medskip}0&\frac{1}{\sqrt{3}}&0&0
\end {array} \right]    
\end{eqnarray}
\begin{eqnarray}
\sigma^{331}&=&\left[ \begin {array}{cccc} 0&\frac{1}{\sqrt{3}}&0&0\\ \noalign{\medskip}
\frac{1}{\sqrt{3}}&0&-\frac{2}{3}&0\\ \noalign{\medskip}0&-\frac{2}{3}&0&\frac{1}{\sqrt{3}}
\\ \noalign{\medskip}0&0&\frac{1}{\sqrt{3}}&0\end {array} \right]  \,,~~ 
\sigma^{332}= \left[ \begin {array}{cccc} 0&-i\frac{1}{\sqrt{3}}&0&0\\ \noalign{\medskip}i
\frac{1}{\sqrt{3}}&0&i\frac{2}{3}&0\\ \noalign{\medskip}0&-i\frac{2}{3}&0&-i\frac{1}{\sqrt{3}}
\\ \noalign{\medskip}0&0&i\frac{1}{\sqrt{3}}&0\end {array} \right]  \,.
\end{eqnarray}
\begin{equation}
	\sigma^{123}= \left[ \begin {array}{cccc} 0&0&-i\frac{1}{\sqrt{3}}&0\\ \noalign{\medskip}0
	&0&0&i\frac{1}{\sqrt{3}}\\ \noalign{\medskip}i\frac{1}{\sqrt{3}}&0&0&0
	\\ \noalign{\medskip}0&-i\frac{1}{\sqrt{3}}&0&0\end {array} \right] 	\,,
\end{equation}
complete the set of $\sigma$-matrices that could generate an interesting new algebra.

%\end{twocolumn}


\begin{thebibliography}{90}
	
\bibitem{RS}
W. Rarita and J. Schwinger,  Phys. Rev. 60 (1942) 61.

\bibitem{VZ}
G. Velo and D. Zwanziger,  Phys.Rev. 186 (1969) 1337.
	
\bibitem{Barut}
A. O. Barut,  I. Muzinich, and D. N. Williams,  Phys. Rev. 130.1 (1963) 442.

\bibitem{JW1}
H, Joos,  Fortsch. Phys., 10:65 (1962) 146. 

\bibitem{JW2}
S. Weinberg, Phys. Rev. 133 (1964) B1318. 

\bibitem{C1}
M. Fierz,  Helv. Phys. Acta, XII, 3 (1939) 37.

\bibitem{C2}
M. Fierz and W. Pauli, Proc. R. Soc. A, 173 (953)  (1939) 211.

\bibitem{C3}
V. Bargmann and E. P. Wigner, Proc. Nat. Acad. Soc.{34}  (1948) 211.

\bibitem{A} 
G. Z. T\' oth, 	Eur. Phys. J. C 73 (2013) 2273.

\bibitem{A1}
E. G. Delgado-Acosta, M. Kirchbach, M. Napsuciale, and S. Rodriguez,
Phys. Rev. D 85 (2012) 116006.


\bibitem{A2}
E. G. Delgado Acosta, V.M. Banda Guzman, M. Kirchbach, 	Eur.Phys.J. A51(3) (2015)  35.

\bibitem{A3}
J. P. Edwards, M. Kirchbach,	Int. J.  Mod. Phys. A  34 (11) (2019) 1950060. 

\bibitem{A4} %gauge
Dario Sauro  Phys.Rev.D 113 (2026)  085015.


\bibitem{B}
J. Escamilla-Mu\^ noz and S. G\' omez-Avila, arXiv:2506.23057.



\bibitem{Ham1}
P. M. Mathews,  Phys. Rev. 143 (1966) 978.

\bibitem{Ham2}
 S. A. Williams, J. P. Draayer and T. A. Weber,  Phys. Rev. 152  (1966) 1207.
 
 \bibitem{N}%Noether for high spin
 N. D.  Minh, arXiv: 0807.4431
	
		
	\bibitem{WKT}
W.-K. Tung,  { Group Theory in Physics}  (World Sci., Philadelphia, 1984).
	
		
	\bibitem{W}
	S. Weinberg, {The Quantum Theory of Fields. Volume 1, Foundations}  (Cambridge Univ. Press, New York, 1995). 
	
	
\bibitem{BDR}
		S. Drell and J. D. Bjorken, { Relativistic Quantum Fields} (Me Graw-Hill Book Co., New York, 1965).
	
\bibitem{Wig}
E. Wigner, {Ann. Math.} { 40}  (1939) 149. 

\bibitem{Mc}
G. Mackey, { Ann. Math.} {55} (1952) 101.

\bibitem{G1}
J. Kiskis, Phys. Rev. D 11 (1975) 2178.


\bibitem{G2}
I. I. Cot\u aescu,  Phys. Rev. D  27 (1983)  2556.







\end{thebibliography}
\end{document}